\documentclass[sn-basic]{sn-jnl}

\usepackage{booktabs}%
\usepackage{array}%
\usepackage{comment}
\usepackage{xcolor}
\setcitestyle{aysep={}}

\jyear{2026}%

\newcommand{\qqbar}{\ensuremath{q\bar q}}
\newcommand{\qqg}{\ensuremath{q\bar q g}}
\newcommand{\JPC}[1]{\ensuremath{J^{PC}=#1}}

\begin{document}

\title[Hybrid Hadrons]{Hybrid Hadrons: Origins and Development}

\author*[]{\fnm{Eric S.} \sur{Swanson}}\email{swansone@pitt.edu}

\affil[]{\orgdiv{Department of Physics and Astronomy}, \orgname{University of Pittsburgh}, \orgaddress{\city{Pittsburgh}, \postcode{15260}, \state{PA}, \country{USA}}}

\abstract{Hybrid hadrons are states in which the gluonic field of the strong force contributes to the structure of the hadron. 
The idea arose in several places at nearly the same time in the mid-1970s because the advent of quantum chromodynamics made it clear that the carriers of the strong force are dynamical degrees of freedom. This resulted in the cavity resonators of the MIT bag model, in the vibrating strings inherited from the dual resonance model, in the Budapest membrane bag, and in a constituent-gluon potential model. The idea was subsequently developed in the early 1980s into a phenomenology capable of predicting masses, quantum numbers, and decay signatures. This article traces that history from the primary literature, giving close attention to dates of submission and to who first proposed what, and supplements the written record with recollections of many of the principal participants. 
It closes by surveying the current state of experimental, theoretical, and lattice research.
}


\maketitle


\begin{quote}
It was as if, suddenly, we had broken into a walled orchard, where protected trees had flourished and all kinds of exotic fruits had ripened in great profusion.\\
\hspace*{\fill}--- C.~F.~Powell~\citeyearpar{Powell:1987}
\end{quote}

\section{Introduction}
\label{sect:I}

The beginnings of the quark model have been thoroughly documented, from its origins in the flavour symmetry ideas of Soichi Sakata~\citeyearpar{Sakata:1956}, and Murray Gell-Mann~\citeyearpar{GellMann:1961} and Yuval Ne'eman~\citeyearpar{Neeman:1961}, to the introduction of quarks as an organising principle of strongly interacting particles (hadrons) by Gell-Mann~\citeyearpar{GellMann:1964} and Zweig~\citeyearpar{Zweig:1964}, to the acceptance of them as dynamical degrees of freedom that build hadrons, much as electrons and nucleons build atoms~\citep{Dalitz:1965}.

The notion of quarks as physical degrees of freedom came as a revelation to Zweig:

\begin{quote}
I remember being very surprised by Figure 1 [see Fig. 1] \ldots There was an enormous peak \ldots right at the edge of phase space. The fact that the $\phi$ decayed predominantly into $K\bar{K}$ and not $\pi\rho$ was totally unintelligible. \ldots Only conservation laws suppress reactions. Here was a reaction that was allowed but did not proceed! I had thought that hadrons probably have constituents and this experiment convinced me that they do, and that they are real. \ldots This was a statement about dynamics which indicated that the constituents were not hypothetical objects carrying the symmetries of the theory, but real objects that moved in space-time from hadron to hadron.\\
\hspace*{\fill}--- G.~Zweig, 1999 recollection quoted by \citet{Fritzsch:2002}
\end{quote}

\begin{figure}[ht]
\centering
\includegraphics[width=0.5\textwidth]{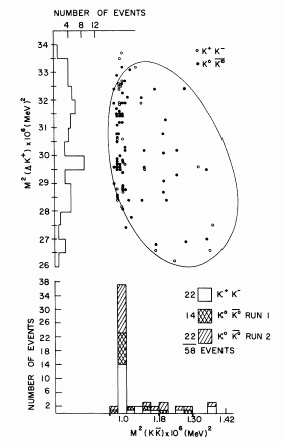}
\caption{Dalitz plot and $K\bar{K}$ invariant-mass distribution showing the $\phi$ peak at the edge of phase space; the figure referred to by Zweig. From \citet{Connolly:1963}. Reproduced with permission.}\label{fig:phi}
\end{figure}

\noindent
Yet, in spite of the experimental motivation, the constituent picture was not well received. Zweig recalled:
\begin{quote}
The reaction of the theoretical physics community to the ace [quark] model was generally not benign. Getting the CERN report published in the form that I wanted was so difficult that I finally gave up trying. When the physics department of a leading university was considering an appointment for me, their senior theorist, one of the most respected spokesmen for all of theoretical physics, blocked the appointment at a faculty meeting by passionately arguing that the ace model was the work of a ``charlatan.''\\
\hspace*{\fill}--- G.~Zweig~\citeyearpar{Zweig:1980}
\end{quote}

A foundational aspect of the quark model is that it describes baryons, such as neutrons and protons, as being primarily composed of three quarks in various quantum states, and mesons as primarily a quark and an antiquark in a quantum state. 
For a quark--antiquark (denoted $q\bar q$) system with orbital angular momentum $L$ and total quark spin $S$, parity and charge conjugation are fixed:
\begin{equation}
P = (-1)^{L+1}, \qquad C = (-1)^{L+S}.
\label{eq:PC}
\end{equation}
These rules forbid the combinations
\begin{equation}
J^{PC} = 0^{--},\ 0^{+-},\ 1^{-+},\ 2^{+-},\ 3^{-+},\ \ldots
\label{eq:JPC}
\end{equation}
which are therefore called \emph{exotic}\footnote{One often sees any particle with unusual content called `exotic', while those of Eq.~\ref{eq:JPC} are called `quantum number exotic'.}. A meson that carries exotic quantum numbers cannot be a simple $\qqbar$ state, and its discovery would be unambiguous evidence for degrees of freedom beyond those of the quark model. 

The advent of quantum chromodynamics (QCD) as the theory of the strong force provided a path to realising this formerly obscure possibility. In particular, the strong force is carried by \emph{gluons}, which are analogues of the photons of quantum electrodynamics. Gluons are degrees of freedom in nature, and their effect must manifest somewhere. Photons appear in enormous numbers, giving rise to nearly classical fields that describe much of our experience of matter. Gluons do not behave in the same way (for reasons that remain poorly understood) --- so they must manifest in other ways. The currently accepted way is that they appear as explicit degrees of freedom in \emph{exotic} states. These exotic states can be thought of as a new form of matter, one that exists for only brief times, but that informs the way the strong force behaves and that may have had an influence on the early universe or on present-day exotic structures such as neutron stars.

In brief, a \emph{hybrid} meson is a quark--antiquark state in which the gluonic field is excited and contributes to the quantum numbers of the hadron; hybrid baryons are defined analogously. The lightest hybrid mesons are expected to include the exotic $\JPC{1^{-+}}$, and the search for this state has been a central goal of light-meson spectroscopy for four decades~\citep{Meyer:2015}.

This article is concerned with the genesis of the hybrid concept: who first conceived of states containing excited glue, how the idea was expressed in the different models of the time, and how it was developed into a testable phenomenology. The story is a case study in near-simultaneous discovery. Within about eighteen months, in 1975--1977, the idea appeared, independently, in at least four guises: as a quark--antiquark--gluon state in a spherical cavity, as a vibration of a string that binds quarks, as an excitation of the gluon field and membrane of a bag with surface tension, and as a three-body bound state of constituent quarks and a massless constituent gluon. Most of these early ideas were set aside, published late, or treated as by-products, and the subject did not acquire its modern shape until the early 1980s, when interest in glueballs created an audience for it. Even its name was contested.

Priority in such a story is best established from the written record, and I have tried throughout to give the dates on which manuscripts were received by journals, or the dates printed on preprints, rather than publication years, which can mislead by a year or more. The written record is, however, incomplete: the motivations of the participants, their interactions, and the fate of ideas that were never published are recorded only in memory. Over the period 2015--2023 I corresponded with many of the principal participants, and their recollections are quoted at length below\footnote{Quotations from correspondence are reproduced with minimal editing: particle symbols and quantum numbers have been typeset, and obvious typographical slips corrected.}. Recollections are, of course, fallible, and where they conflict with the written record I have said so.

The article is organised chronologically. Section~\ref{sect:strings} describes the dual resonance model and the relativistic string, in which vibrational excitations of the object binding quarks were present from the start but were never interpreted as a separate class of hadrons. Section~\ref{sect:QCD} describes how QCD and the MIT bag model made gluons into dynamical constituents. Section~\ref{sect:genesis} contains the core of the story: the independent conceptions of 1975--1977. Section~\ref{sect:BO} discusses the Born--Oppenheimer picture of heavy hybrids that emerged around 1980, Section~\ref{sect:light} the development of light-hybrid phenomenology in 1980--1983, and Section~\ref{sect:name} the contest over the name. Sections~\ref{sect:flux} and~\ref{sect:lattice} describe the flux-tube model and the first lattice gauge theory computations, which supplied a picture of hybrids quite different from the constituent-gluon picture. Section~\ref{sect:modern} describes how modern lattice spectra, Coulomb gauge QCD, and effective field theory have reconciled these pictures, Section~\ref{sect:epilogue} briefly sketches the experimental situation, and Section~\ref{sect:C} offers some conclusions.

\section{Prelude: vibrating strings before QCD}
\label{sect:strings}

\subsection{The dual resonance model}

Between 1968 and about 1974 much of the theoretical effort in strong-interaction physics was devoted to dual resonance models. Veneziano's amplitude~\citep{Veneziano:1968yb} was constructed to have only zero-width resonance poles, linear Regge trajectories, crossing symmetry, and the property, called duality, that the sum over $s$-channel resonances reproduces the $t$-channel Regge exchanges. It was soon realised by Nambu, Nielsen, and Susskind~\citep{Nambu:1970,Susskind:1970} that the spectrum of the model is that of a quantised relativistic string, and Goto and Nambu created the world sheet description~\citep{Goto:1971}. The leading Regge trajectory corresponds to rotation of the string, and the lower ``daughter'' trajectories to its vibrations. As Goddard, Goldstone, Rebbi, and Thorn put it, ``Ever since the early days of dual resonance models, people have attempted to ascribe the excitation spectrum of those models to a vibrating relativistic string''~\citep{GGRT:1973}.

The string picture was extended in many directions: to functional integrals~\citep{Gervais:1971}, to light-cone quantisation and the interacting string~\citep{GGRT:1973,Mandelstam:1973}, and, famously, to the discovery that the Lorentz algebra of the quantised string closes only in 26 space-time dimensions~\citep{GGRT:1973}. The difficulties with quantisation drove much of this work in a formal direction that led ultimately to supergravity and superstrings. A second branch of the community, however, continued to think of the string as a model for real hadrons.

\subsection{Strings with quarks}

For hadron physics the string required quarks. Chodos and Thorn, in a paper received in December 1973, gave the string a finite mass density and noted that ``in contrast to the massless string, the string with finite mass endpoints has vibrational modes of oscillation''~\citep{Chodos:1974}. Nielsen and Olesen had shown in June 1973 that the Abelian Higgs model admits vortex solutions which, for wavelengths long compared with their thickness, behave like the Nambu string~\citep{Nielsen:1973}, so that string-like objects could emerge from a local field theory. Nambu extended this in 1974 to open strings ending on magnetic monopoles, interpreting the monopoles as quarks bound permanently in pairs~\citep{Nambu:1974}. His paper contains two prescient remarks: that ``the finite thickness of the string and the short-range interactions between string elements will distort deep-lying daughter trajectories'', and that ``massive axial-vector gluons can be created by hadrons''.

The most developed model of this kind was the quark-binding and quark-confining string of Tye~\citeyearpar{Tye:1976}, received in December 1975, and its application by Giles and Tye to charmonium, which is discussed in Section~\ref{sect:genesis}. Andreo and Rohrlich solved a string with massive endpoints exactly in 3+1 dimensions~\citep{Andreo:1976}, again obtaining a family of linear trajectories, one for each vibrational mode.

In all of these models, therefore, the string that confined the quarks could vibrate, and its vibrations generated states in addition to those of a quark--antiquark potential model. In this formal sense the string-model ``daughters'' are the ancestors of the flux-tube hybrids discussed in Section~\ref{sect:flux}, a connection that Nathan Isgur and Jack Paton made explicit in 1985~\citep{Isgur:1985a}. But before QCD these excitations were not regarded as a distinct class of hadrons, and they were not associated with gluons. The participants I consulted agree. Julius Kuti wrote:

\begin{quote}
It is possible that hybrids predated our work but I do not know about it, unfortunately. And somehow it seems unlikely to me that the notion of hybrids predated QCD itself.\\
\hspace*{\fill}--- J.~Kuti
\end{quote}

\noindent and Ted Barnes:

\begin{quote}
Regarding ``pre-QCD'' hybrids, none to my knowledge. We mainly thought of this as confining potentials. QCD really made gluons ``real'' \ldots{} However there was so much regular hadron and quark model physics to worry about back then that glue was an afterthought. Also, in the earliest QCD days (1973--4) people were still worried about Faddeev--Popov ghosts, gauge fixing, and all sorts of formal issues; the subject was very highfalutin mathematical, and wasn't yet stable enough to admit intuition about gluons.\\
\hspace*{\fill}--- T.~Barnes
\end{quote}

\section{Gluons become constituents: QCD and the bag}
\label{sect:QCD}

\subsection{Quantum chromodynamics}

By the early 1970s it was widely accepted that the quarks of Gell-Mann and Zweig (Section~\ref{sect:I}) carry a threefold ``color'' degree of freedom, and in 1973 Fritzsch, Gell-Mann, and Leutwyler~\citeyearpar{Fritzsch:1973} set out the advantages of coupling the quarks to a colour-octet Yang--Mills field rather than to a colour-singlet ``gluon''. In their picture ``the gluons are now just as fictitious as the quarks'', confined like the quarks, and they observed that a crude weak-coupling potential proportional to $\sum_{i\neq j} r_{ij}^{-1} C_{iA}C_{jA}$, where the $C_{iA}$ are the colour charges ``of the various quarks, antiquarks, and gluons'', suggests why only colour singlets are deeply bound. Gluons were thereby placed on the same footing as quarks as carriers of colour. The same year saw the discovery of asymptotic freedom~\citep{Gross:1973,Politzer:1973}, and in 1974 Wilson formulated lattice gauge theory and showed that it confines in the strong-coupling limit~\citep{Wilson:1974}; Kogut and Susskind gave the Hamiltonian version~\citep{Kogut:1975}, which would later supply the foundation of the flux-tube model. The discovery of the $J/\psi$ in November 1974 and the success of charmonium potential models gave great impetus to the idea that hadron spectroscopy could be understood with QCD.

\subsection{The bag model}

The idea of quarks confined to a cavity predates QCD. In a paper received on 24 July 1967, P.~N.~Bogoliubov treated the baryon as three quarks moving independently in a self-consistent scalar mean field, by analogy with the nuclear shell model~\citep{Bogoliubov:1968}. He resolved the problem of quark statistics by assuming three triplets of quarks, a form of the colour degree of freedom (he cites O.~W.~Greenberg). Taking the scalar field to be a spherical well that cancels the large free-quark mass inside a radius $r_0$, and letting that mass become infinite, he confined massless Dirac quarks to a sphere and obtained the proton magnetic moment and the ratio of axial to vector couplings to within 12\% and 7\% of experiment.

The bag model was formulated at MIT in the first half of 1974. Chodos, Jaffe, Johnson, Thorn, and Weisskopf (preprint MIT-CTP-387, March 1974) proposed that a hadron is a finite region of space to which its fields are confined, the confinement being enforced by a constant positive energy density associated with the volume of the region~\citep{Chodos:1974a}. If citations are a guide, they were not aware of Bogoliubov's work until later in 1974. They argued that the model gives Regge trajectories that rise without limit and admits only colour-singlet states, and they solved free fields in a one-dimensional bag. Chodos, Jaffe, Johnson, and Thorn (MIT-CTP-417, June 1974) then placed massless Dirac quarks in a static spherical bag, fixed the single parameter by the average of the $N$ and $\Delta$ masses, and computed the magnetic moment, axial charge, and charge radius of the proton~\citep{Chodos:1974b}.

Jaffe recalls the motivation, including the influence of the GGRT string paper (see Section~\ref{sect:strings}):

\begin{quote}
We were certainly aware of GGRT. Not only had Charles [Thorn] just joined our faculty, but also I believe that Jeffrey [Goldstone] \ldots
had had
conversations with Ken [Johnson] sometime after Ken was on sabbatical at SLAC (1971). I
think that GGRT established that an extended geometrical object (in this case, a
relativistic massless string) could be treated as a Hamiltonian dynamical system
and that its spectrum could be computed. We were (foolishly) optimistic that the
bag, although much more complicated, could be quantized using similar operator
methods.\\
\hspace*{\fill}--- R.L.~Jaffe
\end{quote}

He continues:

\begin{quote}
We \ldots 
knew that a region of space time could be imbued with quantum dynamics if the action for the region was proportional to the world-volume swept out by it.  Such was the picture of the string theory of hadrons, which was just then being developed.  In that case, the action was proportional to the world sheet swept out by the string and the constant of proportionality is the ``string tension''.  Relativistic dynamics required physical excitations of the string to be transverse to its extent.  Quark (or other) fields could be defined on the string (I think the formalism was associated with ``Chan--Paton factors'')\footnote{This is correct: the factors were introduced by Hong-Mo Chan and Jack Paton (to reappear in Section~\ref{sect:flux}) as a way to introduce degrees of freedom to strings~\citep{Paton:1969je}.}.  

Similarly a two-dimensional membrane could be defined relativistically if its action was proportional to the tube swept out by the membrane.  The constant of proportionality is the ``surface tension'' and again, a quark field could be constrained to live on the surface of the membrane.  The problem with using either of these pictures as a basis for a model of hadrons is that in addition to excitations of the quarks, there were physical states associated with dynamical excitations of the string or membrane transverse to the spatial domain (in 3+1 dimensions, two transverse d.o.f. for the string and one for the membrane).  

Ken had the brilliant idea that one could define a relativistic action for a three-dimensional blob in 3+1 dimensional space-time.  Then there would be no geometric excitations, but one could define quark (and gluon) fields over the blob whose excitations would correspond to states of the blob-hadron.  It was essential that the geometrical degrees of freedom were not independent dynamical variables, so for example there was no state in which the quarks remained in the ground state configuration and a ``breathing mode'' of the domain was excited.  (This was little appreciated at the time.)  The coefficient of proportionality for the action swept out by the blob had the dimensions of energy per unit volume (the three dimensional analog of the string tension) and we called it the ``Bag Constant'' B.  B was the energy required to open up a domain in which quarks and gluons could live.  So one had a theory consistent with relativity in which quarks and gluons could interact as weakly as one wished but remain forever confined to the bag.\\
\hspace*{\fill}--- R.L.~Jaffe
\end{quote}

The MIT team's understanding progressed rapidly:

\begin{quote}
We soon realized that this picture could be understood as a ``normal region'' in a color-magnetic superconducting vacuum.  In electrodynamics a superconductor expels magnetic fields (the Meissner effect), so a magnetic dipole (or a magnetic monopole, anti-monopole pair) would have to live in a normal region from which the superconducting condensate was expelled.  This was physically reasonable because the intense magnetic fields close to the magnetic charges would drive the superconducting material back to its normal state.  By analogy, if the QCD vacuum were a color magnetic superconductor, the vacuum would expel color-electric fields, forcing any sources of color electric fields to live in a normal region from which the color magnetic superconducting condensate had been expelled.  The difference in energy per unit volume between the color-normal region inside a hadron and the color superconducting vacuum outside a hadron was the fundamental parameter of the theory, again ``B'' --- the Bag Constant.  

In the original bag paper we showed that the boundary conditions on quark and gluon fields at the bag boundary corresponded to local confinement of color-electric flux.  This picture of hadrons as normal regions in a color-superconducting vacuum remains (to me) the most convincing heuristic picture of confinement and is the reason why the ``Bag Model'' remains a useful description of quark/gluon dynamics within hadrons.  \\
\hspace*{\fill}--- R.L.~Jaffe
\end{quote}

In the model, cavity stability is guaranteed by a constant energy density $B$. Quarks obey the free Dirac equation inside the bag, with linear boundary conditions forbidding the flow of colour current across the surface. Gluons obey the free Maxwell equations, and their confinement requires that the colour-electric field have no normal component and the colour-magnetic field no tangential component at the surface. The lowest gluon modes in a spherical cavity of radius $R$ are a transverse-electric (TE, or ``magnetic'') mode with $J^P=1^+$ and energy $2.744/R$, and a transverse-magnetic (TM, or ``electric'') mode with $J^P=1^-$ and energy $4.493/R$. The lowest quark mode has energy $2.043/R$. A TE gluon thus costs only about one-third more energy than a quark, a fact that would make hybrids in the bag model relatively light.

DeGrand, Jaffe, Johnson, and Kiskis~\citeyearpar{DeGrand:1975} computed one-gluon-exchange corrections to the light hadron masses by treating the gluon field as a classical field produced by the quark colour currents, in analogy with classical electrodynamics. At the same time, a related ``SLAC bag'' was developed by Bardeen, Chanowitz, Drell, Weinstein, and Yan in 1975~\citep{Bardeen:1975}. The bag model was well suited to the problem of gluonic excitations because the cavity treats quarks and gluons in exactly the same way: once quarks are placed in cavity modes, it is natural to ask what happens when gluons are placed in cavity modes as well.

The SLAC bag differed from the MIT bag in its starting point. Rather than postulating a boundary and a volume energy, the authors (preprint SLAC-PUB-1490, September 1974) drew inspiration from the Lee--Wick model~\citep{Lee:1974ma} and considered a local field theory of very heavy spin-$\frac12$ quarks coupled to a scalar field, supplemented by a colour vector interaction that makes states of nonzero triality heavy~\citep{Bardeen:1975}. In a variational, tree-level treatment the scalar field changes its value in a region of space where the quarks become light, so that bound states are much lighter than their constituents. The bag thus emerges as a classical solution of the field equations, rather than being imposed, and the quark wave function is concentrated in a thin shell at the surface of the hadron. I note that Tye's quark-confining string (Section~\ref{sect:genesis}) was modelled on both the MIT bag and the SLAC ``bubble''~\citep{Tye:1976}.

\subsection{Glueballs}

What new states does a theory of confined quarks and gluons admit beyond $\qqbar$ and $qqq$? The first answer was ``glueballs''. In 1975 Freund and Nambu~\citeyearpar{Freund:1975} and Fritzsch and Minkowski~\citeyearpar{Fritzsch:1975} argued that bound states of gluons should exist. In a letter received on 27 October 1975, Jaffe and Johnson~\citeyearpar{Jaffe:1976} computed the masses of ``unconventional states of confined quarks and gluons'' in the bag model. They considered three classes of states: gluonic hadrons, ``type I'' multiquark exotics ($qq\bar q\bar q$), and ``type II'' exotics, which are $q\bar q$ or $qqq$ states with quantum numbers ``which cannot be achieved in the `conventional' quark model.'' Their paper, begun at the Aspen Center for Physics, contains the first explicit reference in print to mixed states of quarks and glue:

\begin{quote}
A further class of states: mixed states of quarks and glue, for example $q\bar q$ plus gluons, has yet to be examined.~\citep{Jaffe:1976}
\end{quote}

\noindent Jaffe and Johnson did not pursue the idea. The remark, and its absence of follow-up, is central to the question of priority, and I return to it in Section~\ref{sect:genesis}.

Two other features of the Jaffe--Johnson letter are of interest. First, among the gluonic states it lists a degenerate $(\mathrm{TE})(\mathrm{TM})$ multiplet with $J^{PC}=0^{-+},1^{-+},2^{-+}$ at 1290~MeV, so an exotic $1^{-+}$ two-gluon state appeared in the bag model before any $1^{-+}$ hybrid. (Barnes and Close later argued that the $1^{-+}$ cannot be built from two gluons once centre-of-mass effects are treated correctly~\citep{Barnes:1982c}, and Jaffe, Johnson, and Ryzak reached the same conclusion from an operator analysis~\citep{Jaffe:1986}.) Second, the ``type II'' exotics deserve comment. Jaffe and Johnson observed that in the cavity, $q\bar q$ states with one quark in an $S_{1/2}$ mode and the other in a $P_{3/2}$ mode can be combined into both $C=+$ and $C=-$ states, generating, for example, $0^{+-}$ and $2^{+-}$ nonets. They argued that this doubling is not an artefact of the cavity's violation of translational invariance, since ``it seems natural to expect such states in any theory in which the bound state problem is not strictly a two body problem, that is in any relativistic theory'', and they noted that ``the independent reality of the `bag' plays an important role in this effect.'' But in QCD the degrees of freedom are quarks and gluons, and the bag is a surrogate for the gluonic medium. The extra quantum numbers of the type II states are therefore carried by glue, and in a real sense type II exotics and $\qqbar$-plus-gluon states are the same thing. This intrinsic ambiguity in the bag model --- whether ``extra'' states arise from the bag surface or from gluons --- reappeared later as the problem of ``spurious'' centre-of-mass states, which Chanowitz and Sharpe discussed at length~\citep{Chanowitz:1983a}. Barnes recalls that the question was never resolved:

\begin{quote}
In the early excitement about the bag model there was a lot of discussion about whether the bag was a real, dynamical object or not. \ldots{} Later the MIT people (especially Johnson) seem to have decided that the bag wasn't dynamical, and that the list of $\qqbar$ states in the bag was the same as in ordinary potential models, with the CM excitations removed. Feynman's suggestion (to me) was to just project them out by hand, because they were unphysical and it was only a model \ldots{} My impression was that it was all an unresolved muddle about the quantum mechanical  meaning of classical models.\\
\hspace*{\fill}--- T.~Barnes
\end{quote}

\section{Genesis: four conceptions, 1975--1977}
\label{sect:genesis}

\subsection{Barnes at Caltech: hybrids as virtual states}

Ted Barnes arrived at Caltech as a graduate student in 1971 and worked first with Jon Mathews on quantum field theory in de Sitter space. In 1974 he turned to the bag model:

\begin{quote}
I started working on the bag model in ca.\ 1974 because it was a lot like the E\&M I knew well (cavity resonators + Dirac eqn), and it was quite close to experiment \ldots{} My advisor Jon Mathews insisted I go talk to Feynman, since he was working on this too --- Feynman gave me his notes (his own notation, so mine always differed from MIT) and after 2 weeks invited me to be his grad student so we could work on this. \ldots{} So, I continued working on this with Mathews as an advisor, and published my 1st preprint in April 1975 on the bag.\\
\hspace*{\fill}--- T.~Barnes
\end{quote}

\noindent That preprint, which extended the bag model to massive quarks~\citep{Barnes:1975}, was followed by a key idea:

\begin{quote}
While sitting poolside in Pasadena in Spring/Summer 1975 and thinking about gluonic modes in the bag model, it was clear to me that the MIT-type results for bag model OGE [one gluon exchange] spectroscopy and their big spin-spin splittings were really $\qqbar \leftrightarrow \qqg$ mode mixing, with $\Delta E$ coming in 2nd order $\qqbar \to \qqg \to \qqbar$. So, I worked out the properties of $\qqg$ states (all ``magnetic'' color and spin labels explicit!), and to do $\pi$--$\rho$ etc.\ splitting I found $\qqg$ modes with $0^{-+}$ and $1^{--}$ quantum numbers (I dropped a $1^{-+}$ $\qqg$ state I also found, as being irrelevant, oopsie.), and worked out the $\pi$--$\rho$, $K$--$K^*$ etc.\ splittings. The fitted $\qqbar$--$\qqg$ mixing angles were LARGE, which was disturbing and was ignored. I also did N--$\Delta$ with $qqqg$ intermediates.\\
\hspace*{\fill}--- T.~Barnes
\end{quote}

The idea was thus that the spin-dependent forces responsible for the $\pi$--$\rho$ and $N$--$\Delta$ splittings arise from mixing between the ordinary quark-model states and states containing one transverse gluon in the lowest (TE) cavity mode. The picture is natural in time-ordered perturbation theory, where intermediate states are saturated by eigenstates of the unperturbed Hamiltonian. These quark--gluon states were basis states in a truncated Hilbert space, not, in Barnes's view at the time, hadrons of interest in their own right:

\begin{quote}
My interest in hybrids was as basis states that you needed to calculate OGE spectroscopy. Yes there were extra states that started as $\qqg$ and $qqqg$, but they were pushed up somewhere higher in mass by the mixing, and the mixing angles were pretty big, so everything was all mixed up anyway. I never even thought about the $1^{-+}$ I had constructed in looking for $1^{--}$ $\rho$ mixing states as being anything interesting. It just had the wrong quantum numbers to participate in the $\rho$ mixing problem. In those days we were just trying to see if QCD could explain the known ordinary light hadrons, which needed a big spin-spin force.\\
\hspace*{\fill}--- T.~Barnes
\end{quote}

The first appearance of this work in print was a brief remark in a letter on magnetic dipole decays received on 18 May 1976~\citep{Barnes:1976}. There Barnes listed among the assumptions of the naive SU(6) predictions that ``the amplitude to find the mesons in states containing gluons or other exotic configurations may be neglected'', and reported that ``mixing of the $\lvert q\bar q\rangle$ and lowest-lying $\lvert q\bar q A\rangle$ state to lowest order in the quark--quark gluon coupling does not affect this ratio,'' citing his forthcoming Caltech thesis.

The thesis, \textit{Quarks, Gluons, Bags, and Hadrons}, was submitted on 24 January 1977~\citep{Barnes:1977}. Its chapter on ``Glue Mixing Effects'' (pp.~86--122) states the programme explicitly: ``we shall consider an approach which treats quarks and gluons symmetrically,'' writing a meson as $\lvert m\rangle = a\lvert \qqbar\rangle + b\lvert \qqbar A\rangle + c\lvert \qqbar AA\rangle + \cdots$ and truncating at one gluon. With the lowest TE gluon ($J^P=1^+$) the $\qqbar$ pair in its ground state can form $\lvert \qqbar A\rangle$ states with $J^P=1^-$ (from $\qqbar$ spin $0$) and $J^P=0^-,1^-,2^-$ (from $\qqbar$ spin $1$). Barnes derived the charge-conjugation rule $C\lvert \qqbar A\rangle = (-1)^{\ell+s+1}\lvert \qqbar A\rangle$, and noted that the spin-triplet $J^P=1^-$ state therefore cannot mix with the $\rho$ (eqs.~259--262 of the thesis). That state is the exotic $\JPC{1^{-+}}$, although it is not identified as such. The $\pi$ and $\rho$ were found to contain approximately 36\% and 28\% $\lvert \qqbar A\rangle$ respectively, with mixing angles of $-37^\circ$ and $32^\circ$; Barnes remarked that ``we have driven our model near saturation''.

Less well known is that the thesis also contains what appears to be the first discussion of the experimental signature of a heavy hybrid. Applying the same model to charmonium, Barnes estimated that the predominantly $\lvert c\bar c A\rangle$ vector and pseudoscalar states lie near 4.5 and 4.9~GeV, and argued that the gluon must convert into a colour-octet light-quark pair, each member of which must accompany one of the charm quarks:

\begin{quote}
The signature of these largely $\lvert c\bar cA\rangle$ vector mesons will be (1) strong decays into $(c\bar q)(\bar c q)$ rather than $(c\bar c)(q\bar q)$ \ldots, (2) suppression of the rate to $e^+e^-$ \ldots{} because only the $\lvert c\bar c\rangle$ component may annihilate into a photon, and presumably (3) the existence of a $0^-$ $\lvert c\bar cA\rangle$ state $\sim 500$~MeV above the vector state which will also decay into $D^+D^-, \ldots$ rather than $\chi_c\pi, \ldots$~\citep{Barnes:1977}
\end{quote}

\noindent He proposed the $\psi(4414)$, whose leptonic width was anomalously small, as ``a possible candidate for the $\lvert c\bar cA\rangle$ vector''. This proposal anticipates by almost three decades the discussion of the $Y(4260)$, observed by BaBar in 2005~\citep{Aubert:2005}, as a charmonium hybrid~\citep{Zhu:2005,Close:2005,Kou:2005}.

Barnes's contact with the MIT group bears on the question of priority. Jaffe gave a seminar at Caltech before April 1975, possibly before November 1974; Johnson visited later:

\begin{quote}
I certainly talked to Johnson (1976? actual date uncertain) about gluonic matrix elements and hybrid basis states when he visited Caltech for his seminar, because we discussed color factors. \ldots{} I told him about how I was getting ``spin-spin'' splittings of regular hadrons in the bag model as a mixing problem with q+g states, like $\qqbar$--$\qqg$ and $qqq$--$qqqg$, with explicit hybrid basis states. That's what he rejected as ``too complicated.''\\
\hspace*{\fill}--- T.~Barnes
\end{quote}

Determining the dates of Jaffe's and Johnson's visits to Caltech would be useful in establishing 
the flow of ideas at this time. Unfortunately, I have not been able to locate any records.

The thesis results were written up in Trieste in the summer of 1977 as an ICTP preprint, dated September 1977~\citep{Barnes:1977a}, after Barnes
had taken a post at the University of Kuwait. The preprint states that ``the greatest part of this work was carried out in 1975--6'' at Caltech. Its eqs.~(42)--(44) display the two $\lvert \qqbar A\rangle_\rho$ states and their charge conjugation $C=(-1)^{s+1}$, so the exotic $1^{-+}$ is again implicitly present. The paper was received by \textit{Nuclear Physics B} on 24 October 1977 but, after referee requests for comparison with two different variants of the bag model, appeared only in 1979~\citep{Barnes:1979}\footnote{The published article is headed \textit{Nucl.\ Phys.\ B}~\textbf{152} (1979) 171, and is cited in this way by Tanimoto~\citeyearpar{Tanimoto:1982}. The actual citation is \textit{Nucl.\ Phys.\ B} \textbf{158} (1979) 171.}. Its conclusion contains the advice that ``the search for exotic states of the type $\lvert \qqbar A\rangle$ should be more successful, if heavy quark mesons such as $\lvert b\bar b A\rangle$ are sought.'' Barnes later reflected:

\begin{quote}
In some ways it's frustrating in retrospect, Frank and I could have done all this 5 years earlier. This is the danger of just sitting in your office. \ldots{} Also, we are still waiting for experiment, just 30 years later, so maybe we shouldn't feel bad about the delay in theory.\\
\hspace*{\fill}--- T.~Barnes
\end{quote}

\noindent and

\begin{quote}
A lot of spectroscopy. Quite a lot on hybrids. Very little dynamics or strong decays. If only I had been Feynman's student he might have realized that hybrids were an interesting topic in their own right. Especially exotics. Well, maybe next time.\\
\hspace*{\fill}--- T.~Barnes
\end{quote}

\subsection{Budapest: the bag with surface tension}

Barnes was not the only one to have ideas about constituent gluons in bags. At roughly the same time, and independently, Hasenfratz, Kuti, and Szalay in Budapest developed a bag model in which a membrane under surface tension confines the colour gauge field, and through it, the quarks. Their first account appeared in the proceedings of the 1975 Rencontre de Moriond~\citep{Hasenfratz:1975}\footnote{The Moriond 1975 contribution has not been located. Kuti recalls that it has ``very similar content to the P.~Gn\"adig et al.\ PLB in 1976 including a brief discussion on hybrid excitations''.}. A fuller account by Gn\"adig, Hasenfratz, Kuti, and Szalay was received by \textit{Physics Letters} on 22 March 1976~\citep{Gnadig:1976}. For a fixed quark--antiquark pair at large separation the bag elongates into a cylindrical chromoelectric vortex, producing a linear potential. The authors studied small oscillations of this configuration, finding ``standing `electromagnetic' waves of the gluon field between the two disks like in a cylindrical cavity resonator whose wall is a peculiar membraneous material,'' and noted that ``the membrane may also vibrate when the vortex is excited.'' They concluded:

\begin{quote}
Apart from quark excitations in a potential well, there are also gluon and membrane excitations whose spectrum is not described in a static potential picture. Their description is of great theoretical and experimental interest.~\citep{Gnadig:1976}
\end{quote}

\noindent They also showed that in the strong-coupling limit the massive gluon excitations freeze out and only string-like oscillations remain, so that the model interpolates between the bag and the string. When the Budapest group, joined by Horgan and Richard at CERN, returned to the problem in 1980 (Section~\ref{sect:BO}), they referred to this work as ``our first qualitative prediction'' of $\qqbar$-glue states~\citep{Hasenfratz:1980}.

\subsection{SLAC: vibrations of a charm string}

Inspired by the MIT and SLAC bag models, Henry Tye constructed a ``quark-binding string'' in a paper received on 11 December 1975~\citep{Tye:1976}. The construction confines quarks to a line in space; without quarks it reduces to the massless relativistic string. Because the quark-binding string had unwanted features --- its ground state is an empty string, and it admits states with any number of quark excitations --- Tye also constructed a ``quark-confining string'', in which colour-triplet quarks interact with SU(3) gauge fields but no colour-electric flux may leave the string, so that only colour singlets exist. As Tye put it, ``There are no gluon degrees of freedom; only color electric fields are left.'' Massive quarks sit at the ends of the string, one at each end, ``linked by the appropriate electric flux line, which also defines the string.''~\citep{Tye:1976} This appealing physical picture has remained in the field ever since, with a modern reflection in the flux tube densities computed in lattice field theory~\citep{Bali:1994de}.

Giles and Tye applied Tye's quark-confining string to the newly discovered charmonium spectrum in a letter received on 13 August 1976~\citep{Giles:1976}. The model has no independent gluon degrees of freedom, and hence no glueballs; the colour-electric field is confined to the string. In the Schr\"odinger limit it reduces to charmonium with a linear potential ``plus additional vibrational levels.'' Quantising the vibrational modes semiclassically, Giles and Tye obtained effective potentials $V_n(r)$ for each vibrational level and, with the charm-quark mass and string tension fitted to the $\psi$ and $\psi'$, predicted:

\begin{quote}
The vibrational levels (that are absent in charmonium) start coming in at around 4~GeV. Comparing the wave functions at the origin, we expect the vibrational states to have smaller leptonic widths. \ldots{} We note one prediction at this stage: There are two states around 4.4~GeV in the $e^+e^-$ channel.~\citep{Giles:1976}
\end{quote}

\noindent This is, as far as I can determine, the first published numerical prediction of the masses of what are now called hybrid mesons, and the first proposal that such states might already be visible in the rich structure of $e^+e^-$ annihilation near 4~GeV. A longer paper followed~\citep{Giles:1977}. Isgur and Paton later remarked that the lowest vibrational potential of Giles and Tye ``is similar to ours, but we do not agree with the centrifugal barrier and angular momentum assignments they make''~\citep{Isgur:1985a}.

\subsection{Caltech again: constituent gluons}

The fourth conception arose in the same locale as the first. Horn, visiting Caltech from Tel Aviv, and Mandula, at MIT, submitted ``a model of mesons composed of a quark, an antiquark, and a gluon'' on 28 January 1977~\citep{Horn:1978}. Their motivation is stated plainly:

\begin{quote}
From the analysis of deep-inelastic electron scattering, we know that a major portion of the proton momentum is carried by neutral components which are customarily identified with gluons. In this context, the gluons play a role which is analogous to that of the (current) quarks. In this paper we propose that the gluons may also play a role analogous to the (constituent) quarks in bound-state configurations.~\citep{Horn:1978}
\end{quote}

Horn and Mandula treated the gluon as a massless constituent with $J^{PC}=1^{--}$, bound to heavy (charmed) quarks by linear potentials whose strengths are $9/8$ times the charmonium potential (the colour-octet exchange factor), in a Born--Oppenheimer treatment explicitly modelled on the diatomic molecule. Their abstract and introduction announce that ``some of the new quark--antiquark--gluon bound states have exotic quantum numbers, that is, they cannot couple to a fermion--antifermion system''. The ground level of their spectrum, with all constituents in $S$ waves, has $J^{PC}=0^{++},1^{++},2^{++},1^{+-}$, the same as a $P$-wave $c\bar c$ state. At the next level, with the $c\bar c$ pair in a $P$ wave:

\begin{quote}
Here we find for the first time states whose quantum numbers are such that they cannot be made from a quark and antiquark alone (``exotics of the second kind''). These are the $1^{-+}$ and $0^{--}$ states.~\citep{Horn:1978}
\end{quote}

\noindent They estimated that this level lies only about 300~MeV above the lowest $c\bar cg$ states, proposed that the gluon decays by conversion into a light $\qqbar$ pair, and suggested that the $\psi(4028)$, with its anomalously large $D^*\bar D^*$ branching fraction, might be a $c\bar cg$ vector. They concluded that ``it seems to us reasonable to assume that [the gluon] should play an important role as a spectroscopic constituent also, in addition to being the carrier of the confining forces.''

Barnes, and Horn and Mandula, were working at Caltech at the same time without, it appears, discussing their work in detail:

\begin{quote}
While doing this I learned that Horn and Mandula (also at Caltech, other half of the floor at Downs--Lauritsen) had been interested in $c$-$\bar c$-$g$, and were writing up their results on it! They assumed $J^P(g)=1^-$, which misled them into considering hybrids that were not in the lowest multiplet. My PhD accordingly has a ref to them, but it's a floating number with no text link. \\
\hspace*{\fill}--- T.~Barnes
\end{quote}

\noindent The thesis bibliography indeed lists ``D.~Horn and J.~Mandula, A Model of Mesons with Constituent Gluons, ERDA Research and Development Report (In preparation January 1977)'', and its acknowledgements thank Mandula for his ``interest and suggestions''. The Horn--Mandula paper did not appear until February 1978.

\subsection{Precedence discussion}

Table~\ref{tab:firsts} collects the earliest dated appearance of the principal ideas. The record supports the following conclusions.

\begin{table}[t]
\caption{Earliest dated appearances of key ideas. Dates are those on which manuscripts were received by the journal, or the dates printed on preprints and theses; dates in brackets are recollected.}
\label{tab:firsts}
\begin{tabular*}{\textwidth}{@{\extracolsep\fill}p{0.36\textwidth}p{0.28\textwidth}p{0.26\textwidth}}
\toprule
Idea & Authors & Date \\
\midrule
$\qqg$ and $qqqg$ basis states in the bag; $\qqbar$--$\qqg$ mixing; (implicit) $1^{-+}$ & Barnes & [spring/summer 1975]; thesis 24 Jan.\ 1977 \\
gluon and membrane excitations of a $\qqbar$ bag & Hasenfratz, Kuti, Szalay; Gn\"adig et al. & Moriond 1975; 22 Mar.\ 1976 \\
``mixed states of quarks and glue \ldots{} yet to be examined'' & Jaffe, Johnson & 27 Oct.\ 1975 \\
$\lvert \qqbar A\rangle$ components in a published paper & Barnes & 18 May 1976 \\
first numerical masses (string vibrations in $c\bar c$) & Giles, Tye & 13 Aug.\ 1976 \\
open-flavour decay signature; $\psi(4414)$ as $c\bar c g$ & Barnes & 24 Jan.\ 1977 \\
explicit exotic $J^{PC}$ ($1^{-+}$, $0^{--}$) & Horn, Mandula & 28 Jan.\ 1977 \\
exotic $1^{-+}$ $b\bar b g$ below $B\bar B$ threshold & Hasenfratz, Horgan, Kuti, Richard & 27 Mar.\ 1980 \\
light $\qqg$ spectrum with exotics & de Viron, Weyers & 20 Oct.\ 1980 \\
QCD sum-rule $1^{-+}$ near 1~GeV & Balitsky, Dyakonov, Yung & 8 Jan.\ 1982 \\
``hermaphrodite'' & Close & 1980 (Kupari school) \\
``hybrid'' & Fishbane et al.; Tanimoto & Dec.\ 1981 (unlocated); 23 Mar.\ 1982 \\
``meikton'' & Chanowitz, Sharpe & 1982 (LBL-14865) \\
flux-tube hybrids & Isgur, Paton & 20 Dec.\ 1982 \\
first lattice computation & Griffiths, Michael, Rakow & 10 June 1983 \\
\botrule
\end{tabular*}
\end{table}

First, the conception of quark--gluon states as dynamical objects in QCD belongs to Barnes, who constructed $\qqg$ and $qqqg$ basis states and computed their mixing with ordinary hadrons in 1975. His work circulated only as a thesis and a preprint, and was published only in 1979, so it had little immediate influence. Barnes did not regard the hybrid states as interesting in themselves, and he set aside the exotic $1^{-+}$ state that his formalism produced.

Second, the first published statement that such states exist is the Jaffe--Johnson remark of October 1975, together with the Budapest group's discussion of gluon and membrane excitations at Moriond in 1975 and in March 1976. Neither group developed the idea for several years. Close's view is that the MIT group ``made a passing comment and forgot it''.
The Budapest group, by contrast, returned to the problem in 1980 with a detailed calculation.

Third, the first quantitative predictions of hybrid masses were made by Giles and Tye in August 1976, in a string model without gluons. Their states are the direct ancestors of the flux-tube hybrids.

Fourth, the first explicit statement that quark--antiquark--gluon states can carry exotic quantum numbers, and the first identification of the $1^{-+}$ as such a state, is due to Horn and Mandula, whose paper was received four days after Barnes's thesis was submitted.

It is worth emphasising that all four approaches were independent responses to the same stimulus: QCD had made gluons real, the bag and the string offered tractable models of confinement, and the $\psi$ spectrum offered heavy-quark systems in which new states might be found.

\section{Heavy hybrids and the Born--Oppenheimer picture, 1978--1980}
\label{sect:BO}

The heavy-quark systems that had prompted Giles and Tye, and Horn and Mandula, also motivated the next step. In 1979--1980 Peter Hasenfratz and Julius Kuti, then at CERN, joined Ron Horgan and Jean-Marc Richard to revisit the Budapest bag for heavy quarkonia. Richard recalls the origin of the collaboration:

\begin{quote}
The serious discussions with Julius started when I came back from a meeting at Copenhagen. Ken Johnson gave there a nice lecture on the bag model \ldots{} I was not satisfied by his slides on charmonium, and when discussing with him at the cafeteria, he mentioned the ``Budapest variant'' suited for heavy quarks, and the papers by Kuti and Hasenfratz. At CERN, Peter Hasenfratz was also there, and we quickly decided to revisit their model of charmonium. So far they had a spherical approximation for the bag, and we knew that the linear tail of the potential is associated with a cigar-shaped bag. So, with the help of R.~Horgan, who was already at that time a great expert on computing, among other fields, we repeated the charmonium and bottomonium calculation in this model \ldots{} To be honest, the idea of hybrids was more Julius's and Peter's.\\
\hspace*{\fill}--- J.-M.~Richard
\end{quote}

\noindent and its texture:

\begin{quote}
Back to hybrids, Hasenfratz had some bad time with the algebra of quantum numbers for gluons added to quarks, and made great use of the Quantum Mechanics textbook by Landau and Lifchitz. Horgan was spending most of his days in the basement, close to the main computer. \ldots{} I remember that Julius gave a theory seminar just after the completion of our work, and with his usual kindness and fairness, gave most of the credit to his three collaborators, while he was the real ``glue'' of the team.\\
\hspace*{\fill}--- J.-M.~Richard
\end{quote}

The resulting CERN preprint, dated 27 March 1980 and received by \textit{Physics Letters} on 1 April~\citep{Hasenfratz:1980}, adopted an adiabatic (Born--Oppenheimer) treatment. For fixed heavy-quark positions, the bag shape and the gluon field are found by minimising the energy; the resulting static energy is then the potential in a Schr\"odinger equation for the slowly moving quarks. In addition to the ordinary quarkonia, this procedure generates ``a rich spectrum of $\qqbar$ glue states where the transverse gluon adiabatically occupies the bag around the slowly moving quarks''. These states ``bear some resemblance to diatomic molecules'' and are classified, as in molecular physics, by the projection $\Lambda$ of the gluon angular momentum on the interquark axis. The colour-octet $\qqbar$ pair feels a repulsive Coulomb interaction at short distance, and at large distances a flux tube forms. The abstract makes a bold claim:

\begin{quote}
In $\Upsilon$ particle spectroscopy we predict a narrow ($\qqbar$ glue) state with exotic quantum numbers $J^{PC}=1^{-+}$ below the $B\bar B$ threshold. Its experimental confirmation would be the first direct evidence for coloured glue in the hadron spectrum.~\citep{Hasenfratz:1980}
\end{quote}

\noindent The predicted $1^{-+}$ and $1^{--}$ $b\bar b$-glue states lie at $10.49\pm0.2$~GeV, with gluonic excitations of order 500~MeV, vibrational excitations of 200~MeV, and rotational levels 60~MeV apart: ``the three bands of a molecular spectrum''. 
Hasenfratz et al.\ were aware of the constituent-gluon model of Horn and Mandula, in which ``the gluon is treated \ldots{} as a pointlike particle interacting with the $\qqbar$ pair through an ad hoc linear potential,'' and of Giles and Tye's ``attempt to use a specific string model''.

At the same moment, Buchm\"uller and Tye extended the string picture to the newly discovered $\Upsilon$ system~\citep{Buchmuller:1980}, in a letter received on 13 December 1979. The vibrational states, they wrote, ``are simply the quantized excitations of such modes. They correspond to the vibrations of the string.'' They predicted the lowest $1^{--}$ vibrational $b\bar b$ state at $990\pm90$~MeV above the $\Upsilon$, below open-bottom threshold, with a leptonic width of $0.20\pm0.15$~keV, and they suggested that ``in QCD, one may visualize a vibrational state as a coherent color excitation'' that would de-excite by emitting soft pions. A footnote anticipates the next decade: ``in lattice gauge theories, the energy of vibrational modes is one of the few realistic quantities that can be studied with fixed external color sources.''

The Born--Oppenheimer picture of heavy hybrids --- quarks moving in the adiabatic potential of an excited gluon field --- was thus present in both the bag and the string frameworks by 1980. It was to become the standard framework for lattice studies of heavy hybrids (Section~\ref{sect:lattice}) and, much later, for effective field theories of quarkonium hybrids~\citep{Brambilla:1999xf,Berwein:2015,Oncala:2017hop}.

\section{The light hybrid, 1980--1983}
\label{sect:light}

\subsection{Cavity perturbation theory and the Louvain spectrum}

By 1980 the bag model had acquired a systematic perturbation theory. Following T.~D.~Lee's construction of propagators for fields confined to a rigid cavity, Close and Horgan at CERN developed ``bagged QCD'' and showed that one-gluon exchange in the cavity reproduces the MIT hyperfine splittings~\citep{Close:1980}. In a second paper, received in November 1980, they computed the gluonic admixture in the nucleon, including diagrams in which an external probe acts while the gluon is in transit: ``this is a contribution from a $qqqG$ intermediate state''~\citep{Close:1981}. Maciel and Paton in Oxford used the same methods to compute the $q\to qq\bar q$ dissociation amplitude and found hadronic decay amplitudes too small by a factor of twenty~\citep{Maciel:1982}, a disappointment with perturbative gluons that was to influence Paton's subsequent thinking (Section~\ref{sect:flux}).

Frank Close recalls that these calculations led him to the hybrid idea:

\begin{quote}
I had done some work with Ron Horgan while at CERN where we set up perturbation theory in a bag and showed that OgE [one gluon exchange] reproduced the long computations of the MIT group for hfs [hyperfine splitting]. I spoke about this at a summer school (Dubrovnik?) and included a remark that disturbed me: our approach to the hfs ($N$--$\Delta$ split) had a $qqqg$ intermediate state which implied that hybrid (I called it hermaphrodite, which Squires of Phys Lett [Euan Squires, editor at Physics Letters] took objection to later) could not be far away. I commented that hybrid mesons should exist. But I did nothing with this.\\
\hspace*{\fill}--- F.~E.~Close
\end{quote}

\noindent The school was the 1980 Kupari (Dubrovnik) summer school~\citep{Close:1980a}; Close repeated the remarks at the 1981 EPS conference in Lisbon, in a talk titled ``Glueballs, hermaphrodites and QCD: problems for baryon spectroscopy''.

The first computation of the light $\qqg$ spectrum in cavity perturbation theory was made in Louvain by Fran\c{c}oise de Viron and Jacques Weyers, in a paper received on 20 October 1980~\citep{deViron:1981}. They cite an earlier first approximation by de Viron~\citeyearpar{deViron:1978}, published in 1978. Their abstract states that ``exotic states with $J^{PC}=1^{-+},0^{+-},0^{--}$ are found in the energy range $1.57\le E\le1.93$~GeV'', and their conclusion is emphatic:

\begin{quote}
The important point, however, is the $q\bar q g$ states do exist in the bag approximation to hadrons. Some of them are exotic and lie very low but there is simply no way to get rid of them: to the extent that the bag is a viable approximation to QCD and to the hadron spectrum we are led to expect the discovery of these states and, more generally, the discovery of several families of unconventional mesons!~\citep{deViron:1981}
\end{quote}

\noindent They suggested that the best way to find the exotic $1^{-+}$ might be through radiative transitions from vector mesons such as the $\rho'(1600)$ produced in $e^+e^-$ annihilation.

\subsection{Glueball fever}

The decisive change in the climate came from experiment. Around 1980 the Mark~II and Crystal Ball collaborations at SPEAR observed a prominent pseudoscalar, the $\iota(1440)$, and later a tensor, the $\theta(1640)$, in radiative $J/\psi$ decays, a channel expected to be rich in gluonic states. Chanowitz, who had computed inclusive radiative $\psi$ decay in 1975~\citep{Chanowitz:1975}, proposed in 1981 that the $\iota$ was a glueball~\citep{Chanowitz:1981}. Barnes, Close, and Monaghan showed that the bag model could accommodate the $\iota$ and $\theta$ as glueballs~\citep{Barnes:1982a,Barnes:1982b}. By 1982 the situation was confused enough that Bloom's rapporteur talk at the Paris conference devoted a section to the ``glueball fantasy'' and concluded that ``the question of the experimental verification of gluonic mesons is clouded by the difficulty of the theoretical interpretation''~\citep{Bloom:1982}. The difficulty was that the lightest glueballs were expected to have $J^{PC}=0^{++},0^{-+},2^{++}$ and could therefore mix with, and be confused with, ordinary mesons. A state with exotic quantum numbers would suffer from no such ambiguity, and the hybrid $1^{-+}$ was the obvious candidate.

\subsection{QCD sum rules}

The first prediction of hybrid properties that was based on QCD itself, rather than on a model, came from Leningrad. Balitsky, Dyakonov, and Yung, in a letter received on 8 January 1982~\citep{Balitsky:1982}, analysed the correlator of the gauge-invariant quark--antiquark--gluon current $\bar\psi\gamma^\alpha t^a G^a_{\alpha\mu}\psi$ with the Borel sum-rule methods of the ITEP group. They wrote that ``though the state in question can be easily constructed in the bag model, we do not think its predictions to be trustworthy here,'' and concluded that ``a strong peak should manifest itself in the spectral density of this correlation function at $E\sim1$~GeV, which should be interpreted as a quark--antiquark--gluon resonance.'' The prediction of a $1^{-+}$ hybrid near 1~GeV was quoted widely in the bag-model papers that followed, which took it as independent support for light hybrids. A long version appeared in \textit{Yadernaya Fizika}~\citep{Balitsky:1982a}, and the strange partners were treated later~\citep{Balitsky:1986}.

\subsection{Barnes and Close at Rutherford}

After Kuwait and Trieste, Barnes took a postdoctoral position at Southampton in 1978, where he met Close:

\begin{quote}
Fast forward, I got a postdoc in Southampton in Fall 1978, worked on $\qqbar$ potential models for 2 years, and met Frank when he came to Soton to give a talk. It was on the bag model so we had a lot to discuss, and he helped me into a 5 year postdoc job at RAL [Rutherford Appleton Laboratory] in Fall 1980; we brushed off my old calcs, he found much faster diagrammatic methods to get the same results, and we had a great time writing a series of hybrid hadron papers during 1981--1984 onwards.\\
\hspace*{\fill}--- T.~Barnes
\end{quote}

\noindent Close recalls that the immediate trigger was the Louvain paper:

\begin{quote}
I know that it was a paper by Weyers and de Viron mentioning exotic $1^{-+}$ that played a major role in what happened next. It seemed to me they were wrong (that $1^{-+}$ lies below $0^{-+}$) and talking with Ted confirmed this and we wrote a paper. Then this led to Fran\c{c}oise de Viron \ldots{} visiting RAL just before Christmas. \ldots{} So BCD [Barnes, Close, and de Viron] wrote their paper.\\
\hspace*{\fill}--- F.~E.~Close
\end{quote}

The letter by Barnes and Close, ``A light exotic $\qqg$ hermaphrodite meson?'', was received on 15 April 1982~\citep{Barnes:1982c}. It argued, by analogy with Jaffe's light $q^2\bar q^2$ states, that the strong colour-magnetic attraction between a colour-octet $\qqbar$ pair and a colour-octet gluon pulls the $0^{-+}$ and $1^{-+}$ states down, so that ``$\qqg$ mesons may exist as low as 1~GeV in mass''. The authors described their states as ``hybrid or hermaphrodite states which contain both quarks and gluons'', and stated explicitly their disagreement with de Viron and Weyers, who ``find inverted multiplets, whereas ours exhibit the more conventional pattern $M_{2^-}>M_{1^-}>M_{0^-}$.'' They estimated the nonstrange $1^{-+}$ at about 1300~MeV, and they advocated searches in $\psi\to\gamma X$, in $\psi$ hadronic decays, and in $p\bar p$ annihilation, concluding that ``it may be possible to identify the existence of matter containing constituent gluons more readily by searching for this state than by searching for glueballs.''

The full calculation, with de Viron as co-author, was received by \textit{Nuclear Physics B} on 19 October 1982~\citep{Barnes:1983a}. It computed the four ground-state nonets $J^{PC}=0^{-+},1^{-+},1^{--},2^{-+}$ to $O(\alpha_s)$, including transverse gluon exchange, internal Compton scattering of the gluon on the quarks, and annihilation diagrams. A distinctive prediction was that colour-Coulomb annihilation raises the isoscalar $\omega g$ states by about 100~MeV, making them nearly degenerate with the strange $K^*g$ states, ``perhaps the most clear signal that will enable confirmation of hybrid as against other assignments in non-exotic $J^{PC}$ multiplets.'' The authors discussed production at LEAR and SPEAR, and they cited Barnes's thesis and the 1979 paper, Close's Kupari lectures, Horn and Mandula, and Hasenfratz et al.\ as the origins of the hybrid idea. The baryon sector was treated in a further letter received on 27 November 1982~\citep{Barnes:1983b}, which found that the $qqqg$ ground states form a positive-parity SU(6) 70-plet and that the lightest might lie between the Roper resonance and the $P_{11}(1710)$. That letter contains an unusually candid list of possibilities, the last of which is ``(iv) The concept of hermaphrodite might be nonsense: only future work can help to determine this,'' and it ends by asking what mass a $1^{-+}$ would have in lattice calculations.

\subsection{Chanowitz and Sharpe at Berkeley}

At Lawrence Berkeley Laboratory, Michael Chanowitz and his student Stephen Sharpe had arrived at the same problem by a different route. Chanowitz recalls:

\begin{quote}
As a postdoc at SLAC, I was an author of the SLAC bag paper (PRD 11:1094, 1975 \citealp{Bardeen:1975}) and was very familiar with bag models. \ldots{} Around 1980,  the Mark~I [it was Mark II] discovered a prominent resonance at 1440 in $J/\psi\to\gamma K\bar K\pi$ \ldots{} I dug into the existing data from other experiments and concluded that there was also a pseudoscalar in that mass region and that it was more likely to be the dominant component of the Mark~I signal. Given the two gluon channel, I proposed that it might be a glueball (PRL 46:981, 1981 \citealp{Chanowitz:1981}). \ldots{} So I had gluons on the brain and knew in the bag model that they could naturally be treated in parallel with quarks as hadronic constituents. Being blessed with a great student like Stephen it was natural to investigate together what happens when quarks and gluons are both present as hadronic constituents.\\
\hspace*{\fill}--- M.~Chanowitz
\end{quote}

\noindent Sharpe confirms that the idea was his advisor's:

\begin{quote}
I was working on gluon jets ($\qqg$ 3 jet angular distribution in $e^+e^-$ collisions), and Mike had the idea of using the bag model to study constituent glue. It was definitely his idea to do so \ldots{} So he got me working on the spectrum of glueballs and meiktons and glue-gluino balls. \ldots{} There was some experimental evidence for new states and the question was whether they could be meiktons.\\
\hspace*{\fill}--- S.~R.~Sharpe
\end{quote}

The Chanowitz--Sharpe paper, ``Hybrids: mixed states of quarks and gluons'', was received on 20 September 1982~\citep{Chanowitz:1983a}. Its principal novelty was the inclusion of quark and gluon self-energies, fitted to the light hadrons and to the $\iota(1440)$ interpreted as a glueball. The large gluon self-energy they found raised the predicted hybrid masses by several hundred MeV, into the range 1.2--2.5~GeV. They agreed with Barnes and Close, and not with de Viron and Weyers, that the nonets are ordered $0^{-+},1^{-+},1^{--},2^{-+}$. Two of their phenomenological observations proved durable. First, the dominant hybrid decays occur when the valence gluon creates a $P$-wave $\qqbar$ pair, so that the hybrid decays preferentially into one $S$-wave and one $P$-wave meson; for example, the isovector $1^{-+}$ decays into $\pi D$ or $\pi A_1$ (in modern notation $\pi f_1(1285)$ and $\pi a_1(1260)$). Second, they suggested that a $2^{-+}$ peak at 1850~MeV in the $f\pi$ channel observed by the ACCMOR collaboration, close to the $A_3(1670)$, might signal mixing between a $\qqbar$ $D$-wave state and a hybrid. They wrote: ``If, as we expect, valence gluons exist, then hybrids must also exist.''

The Berkeley and Rutherford groups learned of each other's work while it was in progress and compared calculations. The published Barnes--Close--de Viron paper acknowledges that Chanowitz and Sharpe ``pointed out a sign error in our original calculation of the Z-graph Compton diagrams,'' while Chanowitz and Sharpe adopted the Rutherford group's sign for an annihilation diagram. Sharpe recalls ``visiting Oxford and talking to Ted and Frank (and sorting out some differences in the calculations), playing squash with the energizer bunny (i.e.\ Frank), etc.'' The papers were intended to appear together in \textit{Nuclear Physics B}. Close recalls that the Chanowitz--Sharpe paper appeared ``a volume or so before us (and got lots more citations as a result) yet a check on the dates of submission and receipt confirm that BCD were actually first by a day or so''\footnote{The printed dates do not support this: Chanowitz and Sharpe was received on 20 September 1982 and revised on 26 January 1983; Barnes, Close, and de Viron was received on 19 October 1982 and revised on 3 February 1983. The Barnes--Close letter (received 15 April 1982) does, however, predate both; and it is possible that Close was referring to preprint dates.}.

\subsection{Tanimoto and the first decay calculations}

Morimitsu Tanimoto, a Humboldt fellow at Bielefeld on leave from Ehime University, had been studying the decays of the $\iota(1440)$ as a glueball:

\begin{quote}
In 1980--1981, I was working [on] glueballs since [the] $\iota(1440)$ was observed. Then, I [became] interested in the different property of $\qqg$ hybrid meson and glueball, which was my motivation.\\
\hspace*{\fill}--- M.~Tanimoto
\end{quote}

\noindent His letter ``Decay patterns of $\qqg$ hybrid mesons'' was received on 23 March 1982 and published in October 1982~\citep{Tanimoto:1982}. Following Horn and Mandula, he computed two-body decays induced by the conversion of the constituent gluon into a $\qqbar$ pair, finding typical hadronic widths for the states built on a TM gluon but narrow widths for those built on a TE gluon, and he considered whether the $\iota(1440)$ or $\theta(1640)$ might be hybrids. A second paper, received on 23 August 1982, treated the exotic $1^{-+}$~\citep{Tanimoto:1983}. It contains the observation that the overlap integral for the decay of the $1^{-+}$ into two $S$-wave mesons through the gluon-conversion mechanism vanishes identically, ``independent of the specific form of a wave function'', because ``a magnetic gluon $g(1^+)$ does not convert into an $s$-wave $\qqbar$'', so that this mechanism ``forbids a meson decay $h(1^{-+})\to\qqbar(L=0)+\qqbar(L=0)$''. This is an early form of the selection rule that became central to flux-tube phenomenology (Section~\ref{sect:flux}). Tanimoto's 1982 letter is also notable for its title, and for a footnote on terminology, discussed in Section~\ref{sect:name}.

\subsection{Hybrid baryons}

Hybrid baryons were treated by Barnes and Close~\citeyearpar{Barnes:1983b,Barnes:1983c} and, independently and almost simultaneously, by Golowich, Haqq, and Karl~\citeyearpar{Golowich:1983}. Karl recalls that ``the idea of hybrids was in the air but it certainly was not our invention. I think I heard the idea from Frank Close (but I may be wrong!).'' Close in turn recalls that ``Gabriel Karl had something on hybrid baryons around this time, almost contemporaneous to us.''

\section{What's in a name? Hermaphrodites, meiktons, and hybrids}
\label{sect:name}

By the end of 1982 three names were in competition. Close had introduced ``hermaphrodite'' in 1980:

\begin{quote}
I chose hermaphrodite as it was mixed --- both fermions and boson --- and has established existence in biology. Some referee objected and Squires agreed. I said this was puerile (or words to that effect) explaining that hermaphrodite had an established meaning in science, that ``fission'' had originated in biology. \ldots{} In any event, Squires agreed.\\
\hspace*{\fill}--- F.~E.~Close
\end{quote}

\noindent The term survived in the titles of the Barnes--Close letters of 1982 and 1983 and was adopted by Isgur and Paton, who thanked Barnes and Close for convincing them ``that `hermaphrodite' was acceptable terminology''~\citep{Isgur:1983}.

The Berkeley group preferred an alternative classical coinage. Chanowitz:

\begin{quote}
To name these mixed hadrons we consulted a friend of Stephen's who was a graduate student in classics. She suggested ``meikton'' from the Greek for a mixed thing. It does make an unfortunate anagram which Frank Close took advantage of on several occasions, however the anagram was unintended as far as I was concerned.\\
\hspace*{\fill}--- M.~Chanowitz
\end{quote}

\noindent and Sharpe:

\begin{quote}
This part I can claim credit for, kind of. My girlfriend at the time was a classics grad student so I asked her for some suggestions of greek words for ``mixed thing''. There were a few options, but the one Mike and I liked was meiktons (which just means mixed thing). But, as you know, this was up against hermaphrodites from Frank and Ted, and Nuclear Physics settled on the boring middle ground. Too bad. I still like the good greek name better, but I've moved on!\\
\hspace*{\fill}--- S.~R.~Sharpe
\end{quote}

\noindent The Chanowitz--Sharpe preprint LBL-14865 used ``meikton'', as the Barnes--Close baryon letter makes clear when it refers to ``such `hermaphrodites' or `meiktons' ''~\citep{Barnes:1983b}, and the name appeared in the title of one published paper~\citep{Chanowitz:1983b}. The Liverpool lattice paper of 1983 still speaks of states ``variously named hermaphrodites or meiktons''~\citep{Griffiths:1983}.

Close's account of the resolution is that ``the name `hybrid' became agreed on between us and CS as in our preprints we had used hermaphrodite and CS had used `meikton' --- and as they objected to herm and we to [meikton] we settled on hybrid.'' The published Chanowitz--Sharpe paper, revised in January 1983, is titled ``Hybrids: mixed states of quarks and gluons'' and states ``mixed states with valence structure $\bar qqg$, which we call hybrids''; the Barnes--Close--de Viron paper, revised in February 1983, appeared as ``$Q\bar QG$ hybrid mesons in the MIT bag model''.

The word itself, however, predates this agreement. Barnes and Close had already used ``hybrid or hermaphrodite states'' in the text of their April 1982 letter. More significantly, Tanimoto's letter, received in March 1982 and published in October 1982, has ``hybrid mesons'' in its title, and contains the footnote:

\begin{quote}
The term of ``hybrid meson'' comes from the ``hybrid state'' used by Fishbane et al.~[\textit{Flavor symmetry and glueballs}, NBS preprint NBS 81-0896 (1981)], although Close employed the term of ``hermaphrodites''. We use the term of ``hybrid meson'' in this paper.~\citep{Tanimoto:1982}
\end{quote}

\noindent Tanimoto's 1982 letter thus appears to be the first paper with ``hybrid'' in its title. 
Barnes recalls being told by Sidney Meshkov that Paul Fishbane had introduced the term. The Fishbane--Horn--Karl--Meshkov preprint~\citep{Fishbane:1981} was never published, and neither Karl nor Fishbane could locate a copy. Karl recalls its fate:

\begin{quote}
The history is that we wrote this small piece in Aspen in the summer at the physics institute. It tried to fit the pseudoscalar meson masses and as far as I recall, predict the mass of the eta prime meson using some mixing scheme with parameters in it and we obtained a very good fit to the data, but the fit was very sensitive to the values of parameters \ldots{} My collaborators and I decided to send it to Physics Letters, and it was promptly rejected \ldots{} I no longer recall what the physical picture was being proposed. Somehow gluons came in; but I cannot tell if hybrids were involved.\\
\hspace*{\fill}--- G.~Karl
\end{quote}

\noindent Fishbane's only recollection on the point concerned a different coinage: ``my strongest memory is that PRL did not let us use the term oddball for states with non-standard model quantum numbers''\footnote{Unfortunately, the NIST archives were not able to locate a copy of Fishbane et al., ``Flavor Symmetry and Glueballs'', NBS 81-0896 (1981), so this snippet of the history must remain inferred.}.

By 1985 the matter was settled. Isgur, Kokoski, and Paton wrote of ``states, called hybrids, which are not part of the usual quark model''~\citep{Isgur:1985b}, and Jean-Marc Richard summarises: ``I think that the name `hybrid' was not universal at the beginning, and one used also the name `hermaphrodite', but slowly `hybrid' was adopted.''

\section{Flux tubes, 1982--1987}
\label{sect:flux}

A quite different picture of gluonic excitations was introduced in Oxford at the end of 1982 by Nathan Isgur, on sabbatical from Toronto, and Jack Paton. Paton recalls:

\begin{quote}
Nathan Isgur and I were talking together at the beginning of Nathan's sabbatical visit to Oxford in 1983 [1982]. The linear $\qqbar$ potential was at that time talked about as due to a confined ``QCD flux tube''. At the same time some people were thinking of the possibility of states beyond the quark model consisting of $\qqbar$ + gluon. We were dissatisfied with this because the gluon is a weak coupling concept whereas the QCD flux tube is a strong coupling concept. We noted that the concept of a gluon never arises in the Hamiltonian strong coupling formulation of Kogut and Susskind. So we asked the question ``How do non-quark model states arise in the Kogut Susskind formulation?'' That is how we arrived at the flux tube model. Indeed we did note the similarity of hybrids in the flux tube model to what were called ``daughter states'' in the dual string picture of mesons.\\
\hspace*{\fill}--- J.~Paton
\end{quote}

The published letter, ``A flux tube model for hadrons'', states that ``this work was begun after reading'' the paper of Carlson, Kogut, and Pandharipande~\citeyearpar{Carlson:1983} (received July 1982), which had built a model of baryons on Y-shaped SU(3) flux tubes~\citep{Isgur:1983}. Barnes has raised the question of whether Close, who frequently used Paton's office during visits to Oxford from Rutherford, might also have influenced the turn to hybrids; Paton's recollection is that the flux-tube idea was his own. The Isgur--Paton letter thanks Barnes and Close for conversations; no further evidence on this point is available.

In the strong-coupling limit of Hamiltonian lattice QCD, the degrees of freedom ``are quarks and flux tubes, rather than quarks and gluons.'' Isgur and Paton assumed that this picture survives away from strong coupling, that the flux tube can be treated as a nonrelativistic string with a fixed mass per unit length, and that its modes adjust adiabatically to the motion of the quarks. Ordinary mesons then correspond to quarks moving in the potential of the flux tube's ground state, and hybrids to quarks moving in the potential of an excited flux tube. The lowest excitation is a single transverse phonon, carrying one unit of angular momentum about the interquark axis; the resulting effective radial Hamiltonian is
\begin{equation}
H_1 = -\frac{1}{2\mu}\frac{\partial^2}{\partial r^2} + \frac{\ell(\ell+1)-1}{2\mu r^2} + b r + \frac{\pi}{r} + c + V_{\rm sr}(r),
\label{eq:fluxtube}
\end{equation}
where the centrifugal term reflects the angular momentum carried by the string and $\pi/r$ is the energy of the phonon. Hybrids in this model lose the attractive short-range Coulomb and hyperfine interactions and acquire a centrifugal barrier, so they are heavy: the lowest light hybrids were estimated at 1.8~GeV in the 1983 letter and 1.9~GeV in the full paper~\citep{Isgur:1985a}. Because the phonon can carry either sign of angular momentum, the states come as eight nearly degenerate nonets, $J^{PC}=0^{\mp\pm},1^{\mp\pm},2^{\mp\pm},1^{\pm\pm}$, including the exotic $0^{+-}$, $1^{-+}$, and $2^{+-}$. Isgur and Paton stressed that ``the internal structure of these hermaphroditic states is very different from those of models with constituent gluons and their masses are predicted to be considerably higher.'' The full paper, received in August 1984~\citep{Isgur:1985a}, made the conceptual contrast explicit:

\begin{quote}
One widely adopted approach is to proceed by analogy with the ``constituent quark'' to posit the existence of a ``constituent gluon'' with the quantum numbers ($J^{PC}=1^{--}$, color octet) of a gluon of weak-coupling perturbation theory. We pursue the opposite point of view in this paper, basing our picture on QCD in the strong-coupling regime in which the gluonic degrees of freedom have condensed into collective stringlike flux tubes. \ldots{} it certainly seems to us worth pursuing the implications of the fact that in strong coupling there is no direct analog of the gluon, though there is of the quark.~\citep{Isgur:1985a}
\end{quote}

\noindent The paper also confirms Paton's recollection: ``The vibrational hybrids of Sec.~IIIA correspond to the `daughter states' of the Veneziano model.'' The circle from the dual resonance model to QCD was thus closed. A related result from the string side was J.-F.~Arvis's exact solution of the Nambu string with fixed ends~\citep{Arvis:1983}, which gives the static energies
\begin{equation}
E_N(R) = \left[\sigma^2 R^2 + 2\pi\sigma\left(N - \frac{D-2}{24}\right)\right]^{1/2}
\label{eq:Arvis}
\end{equation}
for string excitation level $N$; Arvis wrote out only the ground state ($N=0$), but the formula for excited levels later became a standard benchmark for lattice computations of excited gluonic potentials.

The model's most influential contribution was to hybrid decays. In a letter received on 28 November 1984, Isgur, Kokoski, and Paton~\citeyearpar{Isgur:1985b} applied a flux-tube-breaking decay mechanism, developed by Rick Kokoski in his Toronto thesis~\citep{Kokoski:1987} and closely related to the $^3P_0$ pair-creation model~\citep{LeYaouanc:1973}, to the exotic hybrids. The title of the letter announced its main result: ``Gluonic excitations of mesons: why they are missing and where to find them''. The lowest hybrids ``decay preferentially to final states with one excited meson [e.g.\ $B(1235)\pi$, $A_2(1320)\pi$, $K^*(1420)K$, $\pi(1300)\pi$, \ldots] rather than to two ground-state mesons (e.g.\ $\pi\pi$, $\rho\pi$, $K^*K$, \ldots)''. The explanation was geometric: ``the relative coordinate of mesons $B$ and $C$ is parallel to $r$ and so cannot absorb the unit of string angular momentum about the $r$ axis''. This ``$S+P$'' selection rule, which had been anticipated in the constituent-gluon framework by Chanowitz and Sharpe and by Tanimoto, now had a simple physical picture and quantitative widths. Barnes's assessment is that the letter

\begin{quote}
had great influence. Before that we really only talked about the spectrum, and handwaved about allowed modes. They had the first widely known serious quantitative decay model, and suggested strange [decay] modes like $\pi f_1$ and $\pi b_1$.\\
\hspace*{\fill}--- T.~Barnes
\end{quote}

\noindent Isgur, Kokoski, and Paton, incidentally, list Barnes's 1977 thesis first among their references to earlier hybrid work. They suggested that diffractive meson production and photoproduction, in which the beam's flux tube might be ``plucked'', would be rich sources of hybrids, and they ended with confidence: ``It is our belief that the guide to meson hybrids that we have provided here, while imperfect, should be sufficient to lead to their discovery.''

The flux-tube model largely displaced the bag model as the standard picture of hybrids for the next fifteen years. The operator analysis of Jaffe, Johnson, and Ryzak~\citeyearpar{Jaffe:1986} offered a third, model-independent perspective: the lowest-dimension operator that can create a hybrid is $\bar q G^{\mu\nu} q$, of dimension five, and the lightest hybrid multiplet is $0^{-+},1^{-+},1^{--},2^{-+}$ if the gluonic ground mode is $1^{+-}$.

Barnes, Close, and Swanson \citeyearpar{Barnes:1995} later solved the flux-tube model numerically with a Hamiltonian Monte Carlo method, dropping the small-oscillation and adiabatic approximations of the earlier analytical treatments. They found that the adiabatic approximation underestimates the hybrid masses, and estimated the lightest hybrids at 1.8--1.9~GeV for $u\bar u$, 2.1--2.2~GeV for $s\bar s$, and 4.1--4.2~GeV for $c\bar c$; the same model reproduced the spin-averaged masses of conventional mesons from $L=0$ to $L=3$. They also noted that the flux-tube decay model forbids spin-singlet hybrids, such as the vector, from decaying into two spin-singlet mesons, which distinguishes hybrid from conventional vector mesons, and argued that charmonium hybrids, which lie below $S+P$ thresholds and should therefore be narrow, are a promising place to search.

\begin{figure}[ht]
\centering
\includegraphics[width=\textwidth]{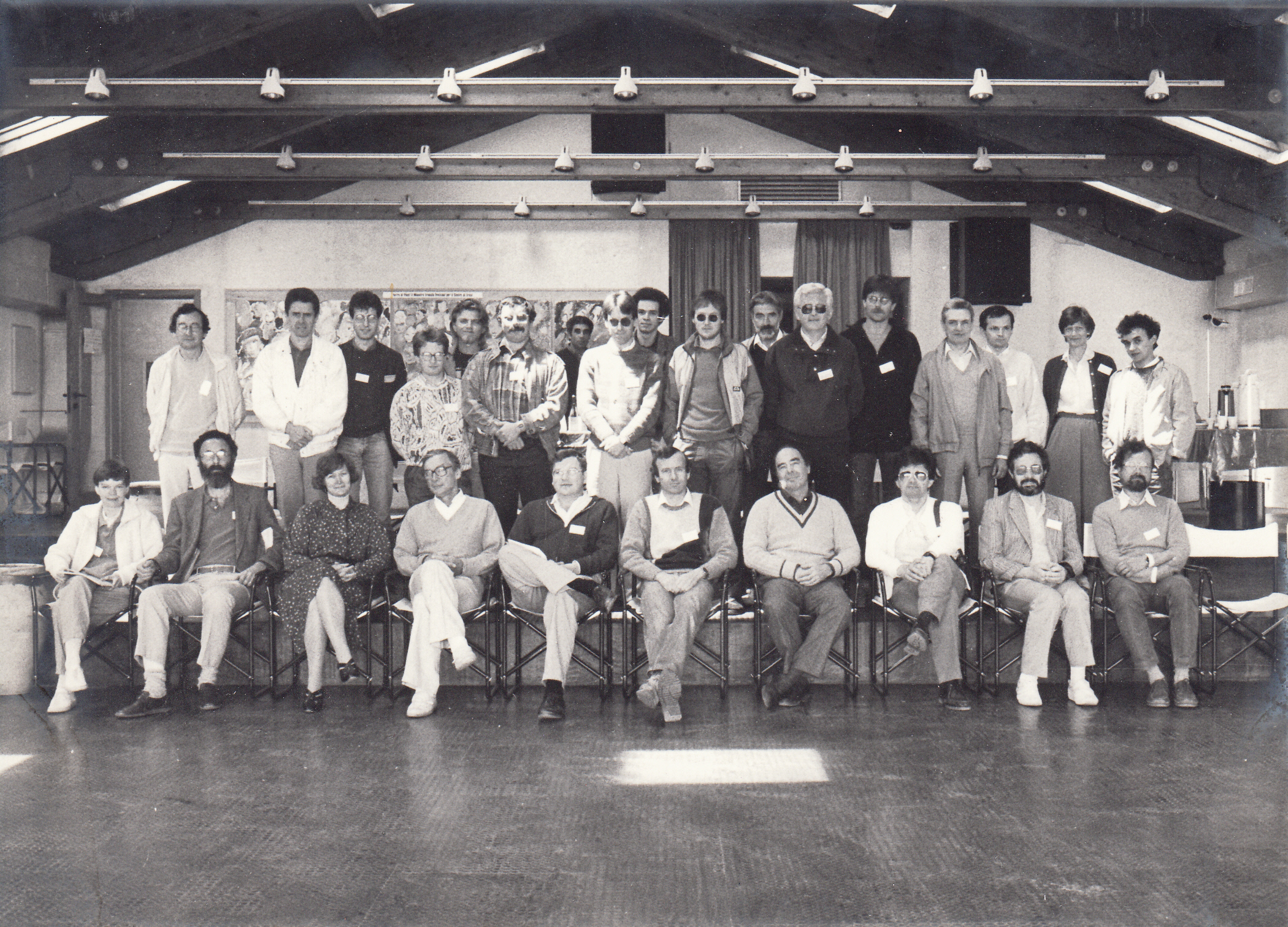}
\caption{Participants at \textsl{Spectroscopy of Light and Heavy Quarks (LEAR)}, 23--31 May, 1987,  Erice, Italy. Front row, second from left: Nathan Isgur; fifth from left: Ted Barnes; sixth from left: Frank Close. Photo credit: Ted Barnes.}\label{fig:erice}
\end{figure}

\section{The lattice, 1983--1999}
\label{sect:lattice}

Lattice gauge theory offered a way to compute gluonic excitations without models. The first such computation was made at Liverpool by L.~A.~Griffiths, Chris Michael, and Paul Rakow, in a letter received on 10 June 1983~\citep{Griffiths:1983}.  Michael describes the background:

\begin{quote}
At Liverpool, we (myself, post-doc Paul Rakow, and PhD students) were using lattice QCD --- gluonic sector --- to make predictions. The glueball spectrum was the obvious target --- but that proved difficult to make advances since we had very limited computing resources. The message from the glueball studies was that (i) a range of different sized operators was important --- and a ``variational'' strategy could separate the ground state from excited states. (ii) it was important to use the lattice symmetries (cubic group for glueballs) to separate the different states. Paul Rakow was key in getting to grips with the discrete group aspect.

Up to that date, potentials between heavy quarks had been extracted using rectangular Wilson loops. This we saw as using a single straight gluonic flux between the sources. We realised that the gluonic flux would ``spread out'' and so used a selection of different gluonic ``paths'' from source to source, with a variational method to separate the ground and excited states. These states were classified by the dihedral group --- so as well as the symmetric ground state, one had excited states (``up-down'' etc) which had excited glue, and non-trivial angular momentum about the interquark axis. This, I realised, was related to hybrid mesons.\\
\hspace*{\fill}--- C.~Michael
\end{quote}

\noindent The computing resources were modest indeed:

\begin{quote}
Most early work (early 1980s) was done on Physics Department IBM Mainframe --- 4341 model --- about 1 mips max \ldots{} We used a discrete subgroup of SU(2) for gluons initially --- since space and time are discrete, why not make the group manifold discrete too. The icosahedral group --- 120 elements --- was a good match for the 3-sphere manifold of SU(2). We stored the group multiplication table --- so a look-up gave the answer to a SU(2) matrix multiply. \ldots{} I later claimed our early processing power was about the same as a modern wrist-watch!\\
\hspace*{\fill}--- C.~Michael
\end{quote}

In more detail, the Liverpool calculation used SU(2) gauge theory on an $8^4$ lattice. Gluonic states between static sources were classified by the lattice symmetry group $D_{4h}$, the cubic analogue of the continuum group $D_{\infty h}$ of a diatomic molecule. The lowest excitation, in the $E_u$ representation (the continuum $\Pi_u$), gives a nearly flat potential, and in the Born--Oppenheimer approximation the corresponding $c\bar c$ and $b\bar b$ hybrids lie about 0.8 and 0.9~GeV above the $J/\psi$ and $\Upsilon$ respectively, with exotic quantum numbers $0^{+-}$, $1^{-+}$, and $2^{+-}$ among them. The authors compared their potential with the bag-model potential of Hasenfratz et al., noting that the Coulomb repulsion of the bag ``is due to fixing the gluon number at one, and so forcing the quark--antiquark pair to be in the adjoint representation,'' and with the Isgur--Paton flux tube, which gave heavier mesons. They concluded:

\begin{quote}
We believe that in the excitations of the gluonic field we have found another part of the hadronic spectrum, in addition to the glueball, which can be largely understood by calculations in the pure gauge sector of the theory, and in which the non-abelian gauge nature of QCD shows up clearly in spectroscopy.~\citep{Griffiths:1983}
\end{quote}

\noindent The calculation was extended to SU(3) in 1984~\citep{Campbell:1984}. Michael's 1985 study of static adjoint (colour-octet) sources~\citep{Michael:1985} addressed the gluonic excitations bound to such a source, now called gluelumps, which govern the short-distance limit of the hybrid potentials. Stavros Perantonis and Michael computed static hybrid potentials in pure SU(3) gauge theory in 1990~\citep{Perantonis:1990}.

Barnes also attempted one of the first lattice calculations of light exotics, but it was never published:

\begin{quote}
In 1984 (1983?) I spent time in Edinburgh and worked on the first LQCD calculations of exotics, especially\ $1^{-+}$ \ldots{} but we only saw noise. I did a lot of debugging but never found the problem. Some people have since said that with contact $\qqg$ operators and such minimal statistics this wasn't surprising. I was leaving for Toronto so this stuff was never published.\\
\hspace*{\fill}--- T.~Barnes
\end{quote}
Amusingly, precisely the same scenario was playing out at nearly the same time across the Atlantic:
\begin{quote}
When I went to Harvard as a Junior Fellow in the Fall of 1983, I was intent on using Lattice QCD to ``do the calculation of hybrids correctly'', i.e. from first principles. I had followed LQCD developments, but had no actual numerical experience, but I met and teamed up with Rajan Gupta (then postdoc at Northeastern) and graduate student Greg Kilcup (officially working with Howard Georgi but in practice mainly with me). Rajan had lattice configurations (quenched, unimproved Wilson fermions, $7^3\times 14$) on a bunch of tapes. We determined the form of the simplest $1^{-+}$ operator (built from $\bar q q$ with an appropriate product of gauge links), coded it up (Fortran?), and, very late one night in the Cambridge computing facility, calculated the two point correlator on the configurations that Rajan had provided (I can't remember how many). This was likely in early 1984, as I remember it being very cold outside. We found ... noise! So we set it aside, and moved on to weak matrix elements.\\
\hspace*{\fill}--- S.~Sharpe
\end{quote}

Calculations of light exotic hybrid masses with propagating quarks in the quenched approximation followed in 1996--1997 from the UKQCD~\citep{Lacock:1996,Lacock:1997} and MILC~\citep{Bernard:1997} collaborations, placing the $1^{-+}$ near 1.9--2.0~GeV and the charmonium $1^{-+}$ hybrid at 4.390(80)~GeV.

The definitive computation of the gluonic excitations of the static potential was carried out by Jimmy Juge, Kuti, and Colin Morningstar using anisotropic lattices~\citep{Juge:1999,Juge:2003}. Morningstar recalls the motivation and a moment of insight:

\begin{quote}
After working on glueballs, excited-gluons in mesons was the next natural step. Christine Davies had told me that their group had tried to get a signal for a hybrid meson and had not succeeded. In fact, she even doubted that a hybrid meson could exist. This certainly motivated me to want to prove it could be done!

We started this work when I was in San Diego. Jimmy Juge was a PhD student there under Julius. Jimmy and I worked closely together to do the coding, runs, and analysis. We used a handful of DEC alpha workstations! For the most part, things went very smoothly. However, there was one channel, the $\Sigma_u^-$, which was a real problem. We just could not observe a plateau in the effective mass for the operators we were using in that channel, suggesting our operators were not creating states with great overlaps onto the low-lying eigenstates. All of our operators were simple paths which connected the static quark and antiquark. I mentioned this to Julius, and he went off to do a quick bag model study of that channel. The bag model could give a qualitative understanding of what the chromoelectric and chromomagnetic fields should look like. When he told me what he found, we realized that ``helical'' paths were needed. When we coded these up and did some runs, we immediately observed the plateaux we were looking for in that channel.\\
\hspace*{\fill}--- C.~Morningstar
\end{quote}

\noindent Kuti, one of the originators of the Budapest bag in 1975, thus used the same model twenty years later to guide the construction of lattice operators. Richard comments:

\begin{quote}
I am not upset when at conferences, young physicists mention ``hybrids, as predicted by lattice QCD'', but I like to stress in private discussions that many concepts were first designed within models and later confirmed by lattice calculations.\\
\hspace*{\fill}--- J.-M.~Richard
\end{quote}

\noindent and Barnes observes that ``LQCD results for exotics and hybrids are now looking rather more like the bag model predictions than the flux tube.''

\section{A modern synthesis: lattice spectra, effective field theory, the quasi-gluon, and Coulomb gauge QCD}
\label{sect:modern}

\subsection{Light hybrids from dynamical lattice QCD}

The quenched lattice calculations of the 1990s established that exotic hybrids exist in QCD, but they could say little about the structure of the hybrid spectrum. That changed after 2009, when the Hadron Spectrum Collaboration combined anisotropic lattices with dynamical quarks, a ``distillation'' method of quark smearing, and very large bases of interpolating operators, including operators built with the chromomagnetic field strength~\citep{Dudek:2009,Dudek:2010}. The resulting light isovector spectrum contains many excited states, including the exotic $0^{+-}$, $1^{-+}$, and $2^{+-}$. Because each state's overlap with each operator could be measured, states with large overlap onto operators containing the gluonic field strength could be identified as hybrids, whatever their quantum numbers.

Jo Dudek then analysed this spectrum~\citep{Dudek:2011}. He found that the hybrid states are organised into ``supermultiplets''. The lightest is $J^{PC}=(0,1,2)^{-+}$ and $1^{--}$, lying roughly 1.2--1.3~GeV above the corresponding $\qqbar$ ground states. This pattern is exactly what is obtained by coupling an $S$-wave colour-octet $\qqbar$ pair to a gluonic excitation with $J^{PC}=1^{+-}$, a chromomagnetic, axial-vector ``quasi-gluon''. The next supermultiplet, with additional exotics, arises when the $\qqbar$ pair is in a $P$ wave. A $1^{+-}$ gluonic excitation with this energy is essentially the lowest TE gluon of the bag model (Section~\ref{sect:QCD}). Dudek's lightest supermultiplet thus coincides with the four ground-state nonets of Barnes, Close, and de Viron and of Chanowitz and Sharpe~\citep{Barnes:1983a,Chanowitz:1983a}, and Barnes's 1975 basis states $\lvert\qqbar A\rangle$ reappear, after three decades, as the organising principle of the lattice hybrid spectrum. The flux-tube model, in which the lowest hybrids form eight nearly degenerate nonets with both signs of $CP$, does not reproduce this splitting into supermultiplets~\citep{Dudek:2011,Meyer:2015}. This is the basis of Barnes's remark, quoted in Section~\ref{sect:lattice}.

Dudek recalled the events that led to this work:

\begin{quote}
I started a PhD in Oxford in 2001 -- in those days new PhD students were expected to search out someone to supervise them pretty rapidly after arriving, as they were supposed to be done with the whole PhD, classes, research, thesis, defense and all, in three years. I really didn't know what I was going to do -- it seemed like everyone around me was doing supersymmetric BSM phenomenology or inflationary cosmology or things too formal for my limited math skills, so I was a bit stuck.

Frank Close had only recently arrived in Oxford from RAL [Rutherford Appleton Laboratory], and for some reason didn't have an office in the theoretical physics building, rather being in with the experimentalists next door. Somehow I ended up talking with him, and he seemed interesting, with his office location reflecting his mindset, being somewhere between theory and experiment. I recall he pitched a few ideas to me: I didn't get far with Generalized Parton Distributions of the proton, made negligible progress with quark-hadron duality, before reproducing some potential model results for radiative transitions in charmonium.

I presented these charmonium results in a short talk at a summer school in Glasgow, where Christine Davies\footnote{Evidently, Davies had an important, if indirect, influence on the field, as this is the third time she has been mentioned.}, the local lattice QCD expert asked ``why don't you just compute these things in lattice QCD?" -- a question which I could do nothing about at the time, but which ultimately led me to Jefferson Lab. I returned to Oxford from this summer school pretty deflated after hearing about all the `fancy' stuff my contemporaries were investigating, but then Frank, who had been spending time at JLab after Nathan Isgur's passing, suggested we look at one of Nathan's last papers applying the flux-tube model to the charge radius of the pion. The idea was that transverse oscillations of the flux-tube would contribute, and that this could be viewed in terms of the photon being absorbed by a quark and `wiggling' the tube, which could excite hybrids. Together with Frank, in my first real calculation, I computed photoproduction rates for hybrids in this model, and then we extended to weak currents to get rates for hybrid production in B-decays, and made some speculative suggestions about hadronic decays.\\
\hspace*{\fill}--- J.~Dudek
\end{quote}
At this point happenstance led him to lattice field theory:



\begin{quote}
I got an offer from Jefferson Lab, who were interested in hybrids because of the GlueX experiment [see Section \ref{sect:epilogue}] that was to be the flagship for their 12 GeV upgrade, and I moved there in the late summer of 2004. At that time the lattice group was David Richards and Robert Edwards, and I suggested to them that we might address Christine Davies' challenge from a couple of years earlier and compute charmonium radiative transitions for the first time in lattice QCD. We could justify this as being related to the GlueX experimental program that was on the way, as once we'd got charmonium working, we would lower the quark mass and think about light meson photoproduction. This project, my first using lattice QCD, was a success, and it led to a relationship with Robert Edwards that has been highly productive and which has brought several other talented people into the field.

In 2008 another student of Frank's at Oxford, Christopher Thomas, came to JLab as a postdoc, and together we developed a basis of fermion-bilinear operators that we hoped to use to study excited mesons. While this wasn't a new idea, mine and Christopher's experience with quark models led us to a set of operators that would end up being somewhat more interpretable than operators that had been used before.

Mike Peardon, who was at JLab on sabbatical, had developed a technique he called `distillation', which made it possible to factorize operator constructions from the quark propagation objects needed to compute correlation functions. This made it practical to utilize our large basis of operators to compute large matrices of correlation functions for various quantum numbers, meaning that we didn't need to make any assumptions about which operators could create which states, we could just let the QCD dynamics decide.


We put all these things together, and analysed with a variational approach (pioneered by others, including Chris Michael) that could extract excited states by virtue of their orthogonality in the basis of operators, leading to an unprecedented level of detail in the determined meson spectrum over many $J^{PC}$.\\
\hspace*{\fill}--- J.~Dudek
\end{quote}

The detailed results then led to the realization that a simple interpretation was possible:


\begin{quote}
Examining the spectrum results in detail, we noticed that operators containing the chromomagnetic construction, but coupled to different gamma matrix structures, leading to non-exotic quantum numbers, dominated the overlap of one particular excited state with $0^{-+}$, one excited state in $1^{--}$ and one excited state in $2^{-+}$, each of which had a mass quite close to the $1^{-+}$ state, and that when we compared the numerical values of the overlaps for each state they were in very close agreement. The implication to us was that the exotic state had non-exotic partners, and a straightforward interpretation was that the color octet of the chromomagnetic construction was neutralizing a color octet from the $q\bar{q}$ pair (in an $S$-wave), and that this was a `supermultiplet' of hybrid mesons\dots
I recall discussing the results with Swanson, and he pushed me to write a separate paper presenting it, and eventually I did, making some comparisons to the older predictions from the bag, flux-tube etc... and pretty definitively killing off my own PhD thesis work on the flux-tube ... ho-hum, never mind.\\
\hspace*{\fill}--- J.~Dudek
\end{quote}

Dudek's picture holds elsewhere in the spectrum. In charmonium the lightest hybrid supermultiplet appears with a similar gluonic excitation energy~\citep{Liu:2012}, as it does in the hybrid baryons, which the lattice finds as positive-parity nucleon and $\Delta$ states above the first band of conventional excitations~\citep{Dudek:2012}. Hybrid decays are now accessible as well. Woss, Dudek, Edwards, Thomas, and Wilson computed the decays of the lightest $1^{-+}$ resonance at an SU(3)-symmetric point~\citep{Woss:2021}. They found a narrow resonance that decays dominantly to $b_1\pi$, which suggests that the $\pi_1$ is broad at physical quark masses. The dominance of an $S$-wave plus $P$-wave final state echoes the selection rules of Tanimoto, Chanowitz and Sharpe, and Isgur, Kokoski, and Paton.

It is worth remarking that the $S+P$ rule has recently been challenged for heavy
hybrids. Using only the symmetries of the Born--Oppenheimer approximation, Roberto
Bruschini showed that the $\Sigma_u^-$ static configuration can couple to a pair of
$S$-wave heavy-light mesons through string breaking, whereas $\Pi_u$ cannot. Hybrids in
the lowest multiplet $H_1$, a bound state of the coupled $\Pi_u$ and $\Sigma_u^-$
potentials, may therefore decay to two $S$-wave mesons, while those in $H_2$ and $H_5$
($\Pi_u$ only) may not~\citep{Bruschini:2024}. The analysis was prompted by a lattice
calculation in which the $1^{-+}$ charmonium hybrid decays to $D^*\bar D$ and
$D^*\bar D^*$ at rates comparable to its decay to $D_1\bar D$~\citep{Shi:2024}. In model language the
created light pair has total spin zero rather than the unit spin assumed since
Tanimoto~\citep{Tanimoto:1982}. 
Although these statements remain contested, future work should help resolve the debate.

\subsection{Effective field theory and hybrids}

The Born--Oppenheimer picture of heavy hybrids, introduced in the bag by Hasenfratz et al.\ and in the string by Buchm\"uller and Tye (Section~\ref{sect:BO}), has since been given a systematic footing in nonrelativistic effective field theory. The starting point is potential NRQCD (pNRQCD), proposed by Antonio Pineda and Joan Soto in Barcelona~\citep{Pineda:1997bj} and developed with Nora Brambilla and Antonio Vairo~\citep{Brambilla:1999xf}, which exploits the hierarchy of scales $m_Q \gg m_Q v \gg m_Q v^2$ in systems of heavy quarks. In the latter paper the authors showed that at interquark distances small compared with $1/\Lambda_{\rm QCD}$ the static hybrid energies are given by the colour-octet potential plus the mass of a gluelump, so that the excited gluonic potentials computed on the lattice by Juge, Kuti, and Morningstar~\citeyearpar{Juge:2003} organise into degenerate multiplets labelled by the gluelump quantum numbers~\citep{Bali:2003jq}. 
Vairo recalls the excitement upon realising that the formalism agreed with the lattice results:

\begin{quote}
While we were writing the 2000 paper [\citealp{Brambilla:1999xf}], Nora and I visited Gunnar Bali in Berlin. I do not remember the details of the discussion we had with him, but it was inspiring enough to make us add to that paper the hybrid/gluelump section. I remember that we were quite excited about the result, as we could explain in an EFT [effective field theory] language the octet short-range behaviour of all hybrid potentials and the degeneracy pattern noticed by Foster and Michael.\\
\hspace*{\fill}--- A.~Vairo
\end{quote}

Braaten, Langmack, and Smith used the Born--Oppenheimer approximation with lattice static energies to compute the hybrid spectra of charmonium and bottomonium and to discuss the $XYZ$ mesons~\citep{Braaten:2014}. Berwein, Brambilla, Tarr\'us Castell\`a, and Vairo derived the corresponding Schr\"odinger equations from pNRQCD~\citep{Berwein:2015}. Because the $\Pi_u$ and $\Sigma_u^-$ energies are degenerate at short distance, the equations couple them, an effect called $\Lambda$-doubling, which lifts the degeneracy of the multiplets $H_1=\{1^{--},(0,1,2)^{-+}\}$ and $H_2=\{1^{++},(0,1,2)^{+-}\}$ and leaves $H_1$ lowest. Oncala and Soto treated spin-dependent effects, decays, and mixing with quarkonium~\citep{Oncala:2017hop}, and Brambilla, Krein, Tarr\'us Castell\`a, and Vairo formulated the Born--Oppenheimer approximation itself as an effective field theory~\citep{Brambilla:2017uyf}. A distinctive result is that spin-dependent hybrid potentials appear already at order $1/m_Q$, whereas in quarkonium they first appear at order $1/m_Q^2$~\citep{Brambilla:2018pyn}. The $H_1$ multiplet is the heavy-quark counterpart of Dudek's lightest supermultiplet; both reflect the $1^{+-}$ quantum numbers of the lowest gluonic excitation, which is the TE gluon of the 1975 bag.

\subsection{Coulomb gauge QCD}

The lattice results raised a conceptual question. Why should a constituent gluon, a weak-coupling concept that Isgur and Paton had rejected in favour of the flux tube, describe the gluonic excitations of a strongly coupled theory? One answer comes from QCD in Coulomb gauge. There the Hamiltonian contains an instantaneous, non-Abelian colour-Coulomb interaction that, in the Gribov--Zwanziger scenario~\citep{Gribov:1977wm,Zwanziger:1998ez}, is confining. Its transverse gluons acquire a large dynamical mass through vacuum condensation and hence can be treated as quasiparticles. Adam Szczepaniak and Eric Swanson developed this framework in a series of papers on the gluonic sector~\citep{Robertson:1999}, on confinement and the constituent representation~\citep{Szczepaniak:2001}, and on hybrids. The first application to hybrids, the heavy hybrids with constituent gluons of \citet{Swanson:1999}, found adiabatic potentials in qualitative agreement with the lattice, but with the wrong level ordering, which was attributed to the neglect of three-body interactions. Decays were treated in \citet{Swanson:1997,Page:1999}. Szczepaniak and Krupinski then showed that a more careful treatment of the non-Abelian Coulomb interaction in the presence of static sources reproduces both the overall scale and the ordering of the lattice gluonic excitations~\citep{Szczepaniak:2006}. Guo, Szczepaniak, and collaborators extended this to the gluelump spectrum, whose level ordering they traced to the non-Abelian part of the Coulomb interaction~\citep{Guo:2008a}, and to the spectrum~\citep{Guo:2008b} and radiative transitions~\citep{Guo:2014} of heavy-quarkonium hybrids. In this picture the quasi-gluon of the lattice is a dressed transverse gluon in its lowest (magnetic, $1^{+-}$) mode, bound to the $\qqbar$ pair by the Coulomb kernel. 
The constituent gluon of 1977 is thus rehabilitated, and the flux tube is reinterpreted as the large-separation configuration of the same dynamics. Recent constituent-gluon models built on this picture address hybrid mixing and decays in detail~\citep{Farina:2020,Swanson:2023,Farina:2024}.

\subsection{Constituent gluons with two degrees of freedom}

A closely related programme was carried out at Mons by Fabien Buisseret, Vincent Mathieu, Claude Semay, Bernard Silvestre-Brac (of LPSC Grenoble), and collaborators. They treated hybrids in a semirelativistic three-body $\qqg$ model with an auxiliary-field technique, compared the flux-tube and three-body descriptions, and computed the $1^{-+}$ masses~\citep{Buisseret:2006}. They then showed that averaging over the gluonic degrees of freedom in the three-body model produces an equivalent excited flux-tube potential~\citep{Buisseret:2007}, which makes explicit the connection between the constituent-gluon and flux-tube pictures that had divided the field since 1983. Mathieu's studies of glueballs and gluelumps argued that the lattice spectra are best described by massless constituent gluons carrying helicity 1, that is, with two rather than three degrees of freedom~\citep{Boulanger:2008,Mathieu:2009b}. His improved calculation of charmonium hybrids with a magnetic gluon supported the interpretation of the $Y(4260)$ as a hybrid meson~\citep{Mathieu:2009a}. This is the identification first suggested, for the $\psi(4414)$, in Barnes's 1977 thesis (Section~\ref{sect:genesis}).

\section{Epilogue: the search for glue}
\label{sect:epilogue}

The theoretical development described above created an experimental programme. Close and Barnes promoted hybrid searches at CERN's Low Energy Antiproton Ring (LEAR), which had been built in part to search for baryonium:

\begin{quote}
Frank sent me to the LEAR workshops in Erice in 1982--3 (?), where I argued that $V_{NN}$ meson exchange models were wrong at short distances (shocked the audience!?), and oh by the way they should also consider looking for hybrid mesons. That eventually took off, and the better (phase 2) LEAR work was those nice spectroscopic Dalitz plots from the Crystal Barrel, with searches for glueballs, hybrids, molecules etc.\\
\hspace*{\fill}--- T.~Barnes
\end{quote}

\noindent Close's view is that ``the modern experimental interest in hybrids arose directly from the CS [Chanowitz--Sharpe] and BCD [Barnes--Close--de Viron] work, which we energetically promoted at LEAR in particular,'' and Barnes notes that in that period ``Europe (LEAR@CERN), Russia (Serpukhov) and the US (BNL, SLAC, some Fermilab) were all making important contributions to light ($udsg$) hadron spectroscopy.''

The first claim of an exotic $1^{-+}$ state, in $\eta\pi^0$, was made by the GAMS collaboration in 1988~\citep{Alde:1988}. Brookhaven experiment E852 reported $1^{-+}$ signals in $\eta\pi$ near 1.4~GeV in 1997~\citep{Thompson:1997} and in $\rho\pi$ near 1.6~GeV in 1998~\citep{Adams:1998}.
The E852 claims were disputed from within the collaboration. A group at Indiana led by Alex Dzierba reanalysed the $\eta\pi^0$ channel and found a $P$ wave, but reported that ``attempts to describe the mass dependence of the amplitude and phase motion with respect to the $D$ wave as a Breit-Wigner resonance are problematic''~\citep{Dzierba:2003fw}. With Szczepaniak they then showed, in a coupled-channel analysis, that known $\qqbar$ resonances ``together with residual soft meson-meson rescattering, saturate the spectra including the exotic $J^{PC}=1^{-+}$ channel'' in $\eta\pi$ and $\eta'\pi$, so that no $\pi_1(1400)$ resonance is needed, while a narrow exotic near 1.6~GeV could not be ruled out~\citep{Szczepaniak:2003vg}. A partial-wave analysis of 2.6 million $\pi^-\pi^-\pi^+$ and 3.0 million $\pi^-\pi^0\pi^0$ events, ten times the sample of the original $\rho\pi$ analysis, found ``no evidence of an exotic meson''~\citep{Dzierba:2005jg}. Meanwhile other E852 analyses reported the $\pi_1(1600)$ in the $f_1\pi$ and $b_1\pi$ channels~\citep{Kuhn:2004,Lu:2005},
 and COMPASS observed the $\pi_1(1600)$ in $3\pi$ in 2010~\citep{Alekseev:2010}. Finally, a coupled-channel analysis later showed that the two E852 states can be described by a single pole~\citep{Rodas:2019}.

In November of 1997, a workshop that Swanson organised, dedicated to the properties of hybrid mesons and their possible discovery modes, was held in Raleigh, NC~\citep{JLabNCSU:1997}.
During the meeting, two participants, Alex Dzierba and Curtis Meyer, decided to mount an effort to find light hybrids at Jefferson Laboratory that was organised in the style of high-energy physics experiments and leveraged a photoproduction
mechanism advocated by Isgur, Kokoski, and Paton in 1985. The effort was eventually successful 
and Jefferson Laboratory now has an additional facility, Hall D, with a dedicated experiment, GlueX, that is currently seeking evidence for hybrid mesons~\citep{Dzierba:2001wg,GlueX:2020idb}. But the road to this point was not smooth, as is often the case when large-scale experimental efforts are mounted.  Meyer recalls:

\begin{quote}
[It started] in 1997 and we got our first physics quality data in 2016, and the first real production run in 2017, close to 20 years! There were so many ups and downs during that time, that much is a blur. We were presented at the 2002 LRP [Long Range Planning] meeting [of the Nuclear Science Advisory Committee (NSAC)], and got into that report as one of the recommendations. Then 2004 hit, and we had an NSAC committee to decide if we should close JLab or RHIC [the Relativistic Heavy Ion Collider at Brookhaven National Laboratory] under tight budget scenarios. We made a tough decision [to close JLab], and then shortly after that things changed and we got CD0 [``Critical Decision 0''---a Department of Energy management milestone]. Things really accelerated then, and by 2007 we were in the midst of the CD2 reviews, that felt like they were almost constant. Then in 2008 we were approved to start construction and in 2009 there was a ground breaking at JLab. The construction was a huge amount of project management, but we were able to install everything by 2013 and in 2014 we had our first commissioning run—amazingly successful!! In 2015, we hoped to have serious physics run, but JLab lost one of the helium plants and we were set back. In 2016 we had our engineering run, and got our first physics quality data. That led to our first publication in 2017 and a slew of Ph. D. theses.\\
\hspace*{\fill}--- C.~Meyer
\end{quote}

In an effort to save money, the GlueX experiment decided to refurbish a superconducting magnet that had originally been built for the Large Aperture Solenoid Spectrometer (LASS) experiment at the Stanford Linear Accelerator Center (SLAC) in 1971. The magnet was subsequently shipped to Los Alamos National Laboratory in New Mexico, where it served the MEGA experiment. After funding approval was obtained, it was brought to the Indiana University Cyclotron Facility (IUCF) for refurbishment in October 2002. This was not an easy task, as the collaboration encountered several problems associated with the original 1970s construction: electrical shorts, leaks in the cooling system, deterioration in the thermal insulation, and failed temperature sensors and strain-gauge instrumentation. This led to future problems for GlueX, although one natural disaster turned out to be fortuitous:

\begin{quote}
As for the magnet, that thing is a tank. It was moved to the IUCF at IU for refurbishing and a great deal of work was done there. I recall several of the iron rings were stored at the edge of the football field parking lot, occupying a couple of spaces. The football team was upset about losing the revenue from those spots, but the iron was too much even for the team to move. It was later moved to JLab to complete the refurbishment and then install it in the hall.

As I recall, the helium level monitoring did not work well and we thought it was fuller than it was. This led to the magnet quenching [a disastrous loss of superconductivity], which it had never done before. The issue was finally resolved when during a hurricane, the lab was looking for places to safely store helium, and they just kept putting it into the magnet. It held a lot of helium! After they had filled it, all the monitoring was consistent and we have never had another quench. Sadly, the operating manuals for magnets a half century old are not always available.\\
\hspace*{\fill}--- C.~Meyer
\end{quote}


\section{Conclusions}
\label{sect:C}

Spurred by the recognition of the role of gluons in QCD, the idea of hybrid hadrons arose almost simultaneously in several places in 1975--1977.
In the MIT bag model it appeared as a by-product of the calculation of hyperfine splittings (Barnes) and as a passing remark (Jaffe and Johnson); in the Budapest bag as gluon and membrane excitations of a colour-electric vortex (Hasenfratz, Kuti, and collaborators); in the quark-confining string as vibrational levels of charmonium (Giles and Tye); and in a potential model as a bound state of quarks and a constituent gluon (Horn and Mandula). The first explicit identification of exotic quantum numbers was made by Horn and Mandula, but the exotic $1^{-+}$ had been constructed, and discarded, by Barnes in 1975.

None of these first conceptions was immediately developed. The thesis work of Barnes was published late and was regarded by its author as an exercise in the spectroscopy of ordinary hadrons; the MIT and Budapest groups turned to other problems; and the vibrational states of strings were overshadowed by the formal development of string theory. The subject was revived around 1980, by the Born--Oppenheimer calculations of heavy hybrids (Hasenfratz, Horgan, Kuti, and Richard; Buchm\"uller and Tye) and by the light-hybrid spectra of cavity perturbation theory (de Viron and Weyers), and it gained an audience because the glueball searches of 1980--1982 had shown how difficult it was to identify gluonic states with ordinary quantum numbers. In 1982--1983 the bag-model calculations of Barnes, Close, and de Viron and of Chanowitz and Sharpe, the QCD sum rules of Balitsky, Dyakonov, and Yung, and the decay calculations of Tanimoto turned the idea into a phenomenology with predicted masses, multiplet structures, production mechanisms, and decay signatures. The names ``hermaphrodite'' and ``meikton'' were superseded by ``hybrid'' as a compromise between the Rutherford and Berkeley groups; the term had apparently been used first by Fishbane and collaborators in an unpublished preprint.

The flux-tube model of Isgur and Paton then replaced the constituent gluon with a collective excitation of the confining string, reconnecting hybrid physics to the dual-resonance-model daughters of a decade earlier, and its decay predictions shaped experimental searches for many years. The first lattice computations, beginning with Griffiths, Michael, and Rakow in 1983, supplied a model-independent picture that, by the late 1990s, had established the existence and structure of gluonic excitations between static sources. Since 2009, lattice spectra have shown that the lightest hybrids form a supermultiplet built on a $1^{+-}$ quasi-gluon. Together with effective field theory and Coulomb gauge QCD, this has largely reconciled the constituent-gluon picture of 1975--1983 with the modern conception of hybrids, while the flux-tube model has fallen away.

Several general lessons emerge. The idea was ``in the air'' --- several of the participants used exactly this phrase --- because a new theory, QCD, met new models and new data at the same moment. The form the idea took depended strongly on the tools at hand: cavity modes produced constituent gluons, strings produced vibrations, and the lattice produced excited flux. The first conceptions were not recognised as important by their authors, and recognition required a change in the experimental situation. Publication delays, rejected manuscripts, and lost preprints shaped the written record considerably, so that the order of publication is an imperfect guide to the order of ideas. Ideas split and merged: bags evolved into dynamical bags, which in turn morphed into flux tubes and strings, bringing the field back to its pre-QCD origins. Meanwhile, the cavity gluons of the bag merged, to some extent, with the constituent gluons of potential models. In the end all the models contained a shadow of the truth --- a truth that is still emerging with more capable lattice field theory computations and an increasingly powerful suite of theoretical methods. 
Finally, the story is not finished: more than four decades after the first predictions, the experimental confirmation of hybrid mesons and the details of their spectrum remain active areas of research.

This work is dedicated to the memories of Nathan Isgur and Ken Johnson, who could not add their reminiscences to this story, and to Jack Paton and Gabriel Karl, who could.

\backmatter

\bmhead{Acknowledgements}

I am grateful to Ted Barnes, Michael Chanowitz, Frank Close, Jo Dudek, Paul Fishbane, Bob Jaffe, Gabriel Karl, Julius Kuti, Curtis Meyer, Chris Michael, Colin Morningstar, Jack Paton, Jean-Marc Richard, Stephen Sharpe, Morimitsu Tanimoto, and Antonio Vairo for sharing their recollections and for their contributions to science.

\section*{Statements and Declarations}

The author has no relevant financial or non-financial interests to disclose.
No funding was received for conducting this study.

\bibliography{HybridHistoryv4}

\end{document}